\documentclass[journal]{IEEEtran}
\usepackage{lipsum}
\usepackage[utf8]{inputenc}
\usepackage{amsmath,amsfonts,amssymb,bm}
\usepackage{graphicx}
\usepackage{subcaption}
\usepackage{caption}
\usepackage{xcolor}
\usepackage{mathtools}
\usepackage{comment}
\usepackage{textcomp}

\usepackage{multirow}
\usepackage{epstopdf}
\usepackage{cite}
\usepackage{arydshln}
\usepackage{flushend}
\usepackage{blkarray, multirow}
\usepackage{arydshln}

\newcommand{\bi}{\begin{itemize}}
\newcommand{\ei}{\end{itemize}}

\newcommand{\be}{\begin{enumerate}}
\newcommand{\ee}{\end{enumerate}}
\newcommand{\bd}{\begin{description}}
\newcommand{\ed}{\end{description}}
\newcommand{\bc}{\begin{center}}
\newcommand{\ec}{\end{center}}
\newcommand{\bt}{\begin{tabbing}}
\newcommand{\et}{\end{tabbing}}
\newcommand{\bfig}{\begin{figure}}
\newcommand{\efig}{\end{figure}}
\newcommand{\beq}{\begin{equation}}
\newcommand{\beqarr}{\begin{eqnarray}}
\newcommand{\beqarrn}{\begin{eqnarray*}}
\newcommand{\eeq}{\end{equation}}
\newcommand{\eeqarr}{\end{eqnarray}}
\newcommand{\eeqarrn}{\end{eqnarray*}}
\newcommand{\bflr}{\begin{flushright}\vspace{-0.2in}}
\newcommand{\eflr}{\end{flushright}}
\newcommand{\bsub}{\begin{subequations}}
\newcommand{\esub}{\end{subequations}}
\newcommand{\barr}{\begin{array}}
\newcommand{\earr}{\end{array}}
\newcommand{\nn}{\nonumber}

\def\undb#1{\mbox{\bf{#1}}}

\def\dn{\stackrel{\scriptscriptstyle \triangle}{=}}

\def\BibTeX{{\rm B\kern-.05em{\sc i\kern-.025em b}\kern-.08em
		T\kern-.1667em\lower.7ex\hbox{E}\kern-.125emX}}

\begin{document}

\title{\huge{MIMO FSO Systems in Hybrid Quantum Noise Environments: SKR Analysis with One- and Two-way CV-QKD Protocols}}
\author{Sushil Kumar, Soumya~P.~Dash,~\IEEEmembership{Senior Member,~IEEE}, and George C. Alexandropoulos, \IEEEmembership{Senior Member, IEEE}

\thanks{S. Kumar and S. P. Dash are with the School of Electrical and Computer Sciences, Indian Institute of Technology Bhubaneswar, Argul, Khordha, 752050 India, (e-mails: \{a24ec09010, spdash\}@iitbbs.ac.in).}
\thanks{G. C. Alexandropoulos is with the Department of Informatics and Telecommunications, National and Kapodistrian University of Athens, Panepistimiopolis Ilissia, 16122 Athens, Greece and also with the Department of Electrical and Computer Engineering, University of Illinois Chicago, IL 60601, USA (e-mail: alexandg@di.uoa.gr).}
}
\maketitle

\begin{abstract}
This paper studies a multiple-input multiple-output (MIMO) free-space optical (FSO) communication system employing continuous-variable quantum key distribution (CV-QKD), with the goal to support secret key transmission between two legitimate users, Alice and Bob. All involved wireless channels are subjected to atmospheric turbulence leading to beam spreading, pointing error, and turbulence-induced fading, which along with the presence of hybrid quantum noise negatively impact secret key exchange. Furthermore, the legitimate MIMO FSO system faces the threat of compromise from an eavesdropper, Eve, employing a collective Gaussian attack to intercept the secret key exchange. Novel one- and two-way protocols for enhancing the security of the transmitted keys are proposed. To this end, the transmissivity of the FSO channels is mathematically formulated and bounds on the mutual information between the transmitted and received coherent states are obtained, which are then used for deriving novel expressions for the secret key rates (SKRs) for both one- and two-way protocols. The presented numerical results corroborate the proposed analytical secrecy framework, quantifying the SKR gains obtained by employing MIMO and the two-way protocol for FSO CV-QKD systems. 
\end{abstract}
\section{Introduction}
The upcoming sixth-generation (6G) communication networks are expected to provide seamless connectivity anytime and anywhere~\cite{9144301, 9145564}. This goal motivates the integration of non-terrestrial networks (NTNs), encompassing satellite-based systems and high-altitude platforms, with terrestrial infrastructures, which is lately receiving substantially research and development interests~\cite{9617565, 10816483, 9992172, 9861699}. Free-space optical (FSO) communication systems constitute the most prominent enabling technology to support such networks, due to their ability to deliver high-capacity, interference-resistant, and license-free links with rapid deployment over extensive distances~\cite{FSO_elamassie_2023_10111210, FSO6g_bekkali2023_10325456,8920091}. However, their deployment over multiple operators and in dynamic topologies inherently leads to the demand for secure data transmissions in the presence of eavesdroppers that compromise legitimate data traffic integrity.

Several studies in the literature have explored the use of classical cryptographic techniques to enhance the security of data transmission in FSO-based systems \cite{cryp_fso, 9944674, 10411843, 6775279}. The available algorithms base their strength in extensive computational complexity, making them practically infeasible to break using conventional computing resources within reasonable time frames. However, with the rapid advancements in quantum computing, particularly through Shor’s and Grover’s algorithms, encryption schemes, such as Rivest–Shamir–Adleman (RSA) and elliptic curve cryptography (ECC), are expected to become vulnerable, as their decryption can be achieved in significantly shorter time~\cite{11175376,Gitonga2025QuantumImpact}. This impending threat has motivated the exploration of alternative approaches for achieving unconditional security, among which quantum key distribution (QKD) has emerged as a promising solution \cite{weedbrook2012gaussian,8732438}. This scheme exploits fundamental principles of quantum mechanics, namely the no-cloning theorem and measurement disturbance, to enable two parties to securely exchange encryption keys, ensuring a level of security that remains impervious to potential advances in computational power \cite{weedbrook_gaoussian_van_numen,neel_spd_THZ_QKD_oct2021,weedbrook2010quantum, pirandola2021limits}.

Among the different categories of QKD protocols, continuous-variable QKD (CV-QKD), which encodes information onto the quadratures of coherent or squeezed states, presents as a viable technology for FSO systems \cite{cv_dv_qkd_Scarani_2009,cvqkd_ppf_2024,sahu_cvqkd_mimo_aug2023}, owing to its advantages including compatibility with standard optical telecommunication components, potentially higher secret key rates (SKR), and improved resilience to photon loss~\cite{pirandola2020advances}. In~\cite{hui_zhao_FSOQKD_jan2021_9220903}, the authors analyzed the transmission characteristics of QKD over a single-input single-output (SISO) FSO channel, focusing on the impact of atmospheric turbulence on the overall transmission probability. Extending this line of research, \cite{minh_multiuse_aug2023_10180066} examined the feasibility of distributing CV-QKD-encoded secret keys from a satellite platform to multiple legitimate ground users, thereby addressing the scalability of QKD for satellite networks. Furthermore, \cite{nancy_uavFSO_jun2022_9747979} investigated the achievable SKR for a SISO FSO link assisted by an unmanned aerial vehicle, where CV-QKD was employed to enhance secure communications in dynamic airborne environments. To mitigate the fractional loss of secret keys caused by the lengthy and complex reconciliation process in one-way QKD, a two-way QKD protocol, leveraging bidirectional communication between the legitimate transceiver pair, was introduced in \cite{Pirandola_2008} and later applied in \cite{Twoway_phuc_2017_7996809} to enhance the secrecy performance in SISO FSO systems.

Most current FSO QKD studies focus on SISO configurations, whose performance is limited by atmospheric disturbances that can deflect or spread the beam, sometimes causing it to miss the receiver's aperture entirely, resulting in connection failure. In addition to these propagation impairments, secret key transmission in FSO systems is further degraded by quantum noise \cite{mouli_hybridnoise_Jan2025}. Such noise originates from environmental interactions, qubit coupling, and imperfect system controls, leading to dephasing and decoherence. Alongside classical noise, quantum noise can significantly degrade quantum channel fidelity and compromise communication reliability. 

To mitigate the adverse effects of channel propagation and fading, multiple-input multiple-output (MIMO) techniques have been recently integrated with CV-QKD, demonstrating notable improvements in the system's performance~\cite{neel_spd_THZ_QKD_oct2021, kumar2024risassistedmimocvqkdthz,sahu_cvqkd_mimo_aug2023,11152330}. Furthermore, it was shown in~\cite{11129674} that employing two-way communication protocols in MIMO FSO systems can further enhance SKR performance. However, existing studies predominantly assume simplified noise models, limiting their applicability to practical FSO scenarios. Moreover, \cite{11129674} showcased that implementing the two-way communication protocol for MIMO FSO systems improves their SKR performance. 

However, to the best of our knowledge, the performance of one- or two-way protocols for such MIMO FSO systems subject to hybrid quantum noise has not been analyzed. In fact, MIMO-assisted CV-QKD under the two-way protocol remains significantly under-explored. Specifically, most studies tend to overlook the simultaneous effects of discrete photon noise and continuous thermal noise. This coexistence is unavoidable in practical FSO receivers operating in the presence of background radiation and imperfect detection, and the presence of hybrid quantum noise is expected to directly impact mutual information and SKR in MIMO systems, where noise accumulation and spatial correlations can significantly affect system performance \cite{4786505}. Motivated by these limitations, this paper addresses these research gaps by considering a MIMO FSO system in which two legitimate users, Alice and Bob, attempt to achieve secure secret key exchange in the presence of an eavesdropper, Eve, who compromises the system's security by employing a collective Gaussian attack. The FSO channels are considered to degrade the secret-key transmission achieved via CV-QKD due to atmospheric turbulence, which causes diffraction, spreading, and misalignment of the transmitted beams, and due to the presence of hybrid quantum noise. The main contributions of this paper are summarized as follows:
\begin{itemize}
\item By accounting for the beam spreading occurring during transmission, which is modeled by considering the field distribution of the beams at the transceiver pair, the pointing error of the beams, which is modeled by the Weibull distribution, and the turbulence-induced fading, modeled by the lognormal distribution, a mathematical framework for the transmissivity of the considered MIMO FSO configuration is presented.
\item A one- and a two-way protocols for the considered MIMO FSO CV-QKD system when operating under hybrid quantum noise, modeled by a Gaussian mixture distribution guided by a Poisson process, are proposed.
\item Novel analytical expressions for the SKR performance of the considered system with both presented protocols and for the case where Bob's receiver utilizes homodyne detection along with reverse reconciliation (RR), while the eavesdropper employs collective Gaussian attack for data decryption, are presented. 
\item The superiority of the two-way protocol over its one-way counterpart, along with the SKR enhancement achieved through the MIMO configuration with respect to SISO, is showcased through an asymptotic analysis.
\end{itemize}

The rest of the paper is organized as follows. Section II presents the MIMO FSO CV-QKD system model, detailing the channel model, the analytical framework for channel transmissivity, and the additive hybrid quantum noise model. The transmission and reception of the secret keys, along with the effect of the collective Gaussian attack by the eavesdropper, are described for the one- and the two-way protocols in Section III. Capitalizing on the statistics of the hybrid quantum noise, Section IV proposes bounds on the mutual information between the transmitted and received states for both considered protocols, which are then utilized, in Section V, to derive closed-form and asymptotic expressions for the SKR performance. The numerical results corroborating the analytical findings and showcasing the dependency of the system performance on the considered parameters are presented in Section VI. Finally, Section VII presents the concluding remarks of the paper.

\textit{Notation:} $\textbf{A}^\dagger$ and $\textbf{A}^T$ denote the conjugate transpose and transpose of a matrix $\textbf{A}$, respectively. $\boldsymbol{1}_{N}$, $\boldsymbol{0}_{N}$, and $\undb{I}_N$ represent a $1 \times N$ vector of ones, the $N \times N$ zero matrix, and the $N \times N$ identity matrix, respectively. $J_0(\cdot)$ denotes the zero-order Bessel function of the first kind, $\jmath\triangleq\sqrt{-1}$, and $\text{diag}(\boldsymbol{a})$ constructs an $M \times M$ diagonal matrix with the elements of vector $\boldsymbol{a}$ along its principal diagonal. We also use the definition $\undb{Z}\triangleq\text{diag}(1,-1)$. The operator $\otimes$ represents the Kronecker product, while $\undb{E}[ \cdot ]$ and $\langle X \cdot Y\rangle$ denote the expectation operator and the quantum correlation between $X$ and $Y$, respectively. Finally, $\hat{Q}$ denotes an operator (such as annihilation or creation) acting on the signal mode $Q$. Furthermore, $\mathcal{N}(\mu, \sigma^2)$ denotes a real Gaussian random variable with a mean value of $\mu$ and variance of $\sigma^2$.
\section{System Model}
\begin{figure*}[!t]
    \centering
    \includegraphics[width=16cm,height=7cm]{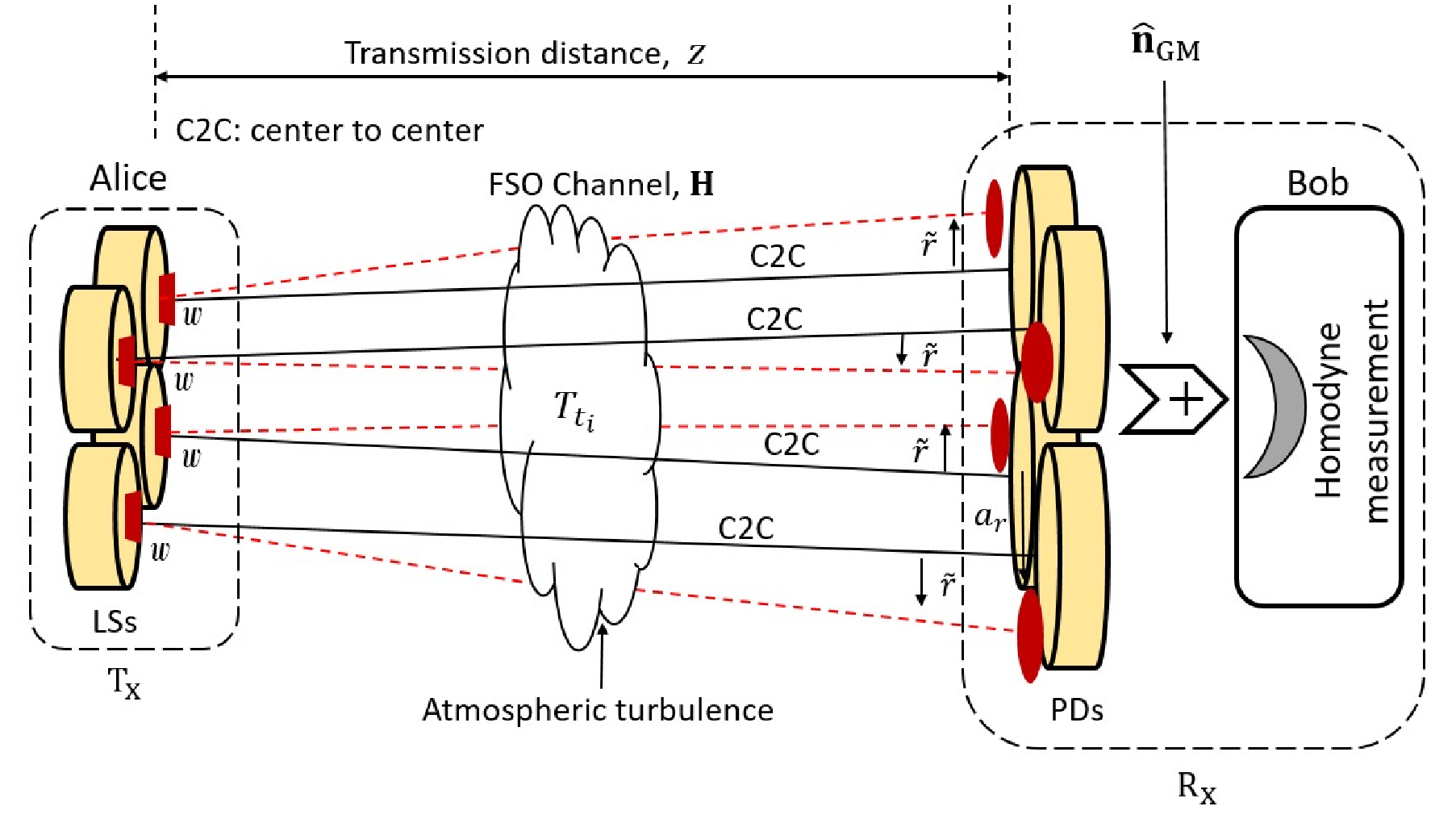}
    \caption{The considered MIMO FSO CV-QKD system model.}
    \label{f1}
\end{figure*}

The considered CV-QKD system, illustrated in Fig.~\ref{f1}, consists of a MIMO FSO channel between two legitimate users, namely, Alice and Bob, equipped with $N_T$ and $N_R$ optical transceivers (or sub-apertures), respectively. Typically, laser sources (LSs) are employed for the transmission phase, and photodetectors (PDs) are used for the detection process. Alice and Bob aim to establish a successful exchange of secret keys using the CV-QKD scheme in the presence of an eavesdropper, Eve, who is targeting to steal the secure information.
\subsection{Channel Model}
The FSO channels are subjected to atmospheric turbulence that leads to diffraction, spreading, and misalignment of the transmitted beams. Thus, we denote the MIMO FSO channel gain matrix by $\mathbf{H} \in \mathbb{C}^{N_R \times N_T}$, with the spatial sub-channel between the $j$-th transmit aperture and the $l$-th receive aperture, with $j \in \{1,2,\ldots, N_T\}$ and $l \in \{1,2,\ldots, N_R\}$, denoted by a complex channel gain $h_{l,j}$, which is modeled as~\cite{Zhao_multiplexFSO_dec2015}:
\beq
h_{l,j} \triangleq \frac{\int_{{\mathcal{D}}_l} G_j \left( r - \tilde{r} \right) \text{d}s }{\sqrt{2\pi \int_{0}^{R_0} r
\left| E_j(r) \right|^2 \text{d}r}}  \, ,
\label{eq1}
\eeq
where ${\mathcal{D}}_l$ covers the receiver's reception area, $E_j(r)$ is the field distribution at the transmitting aperture, expressed as:
\beq
E_j(r)\triangleq\sqrt{\frac{2}{\pi w^2}}\, e^{-\frac{r^2}{w^2}} \, ,
\label{eq2}
\eeq
and $G_j(r)$ denotes the field reaching to $l$-th receiver aperture, which is given as follows:
\beq
G_j(r) = 2\pi \int_{0}^{\rho_{\text{max}}} \rho F_j(\rho) J_0(2\pi r \rho)e^{\jmath \sqrt{k^2-\left(2\pi\rho\right)^2} z} \text{d} \rho\, .
\label{eq3}
\eeq
In the latter expression, $R_0\triangleq\sqrt{N_T} w$ is the radius of the transmitter aperture, $w$ is the waist size of the Gaussian beam at the transmitter's sub-apertures, $k \triangleq2\pi/\lambda$, $\rho_{\text{max}} \triangleq \sin\left(\frac{\lambda}{\pi w}\right)/\lambda$, $a_r$ is the radius of a PD's lens, $\lambda$ is the wavelength, and $z$ is the transmission distance between Alice and Bob. Furthermore, $F_j(\rho)$ in \eqref{eq3} outputs the spatial frequency spectrum of $E_j(r)$, and is obtained as follows:
\beq
F_j(\rho) \triangleq 2\pi \int_{0}^{R_{0}} r \, E_j(r) \, J_{0}\!\left( 2\pi r \rho \right) \, \text{d}r \, .
\label{eq4}
\eeq

It is noted that in \eqref{eq1} the field reaching the aperture of the receiver can be expressed as $G_j \left( r-\tilde{r}\right)$, with the term $\tilde{r}$ arising due to the misalignment factor of the FSO system. As shown in Fig.~\ref{f1}, misalignment can occur due to the following two key factors: \textit{i}) pointing errors, which may arise from mechanical vibrations or tracking imperfections; and \textit{ii}) beam wandering, resulting from variations in atmospheric conditions. Both these factors determine the equivalent radial standard deviation of the beam centroid displacement, $\sigma_{\tilde{r}}$. Assuming that these factors are independent of each other, $\sigma_{\tilde{r}}$ is computed as:
\beq
\sigma_{\tilde{r}} \triangleq \sqrt{\sigma_p^2 + \sigma_{\text{TB}}^2} \, ,
\label{eq5}
\eeq
where $\sigma_p^2 \triangleq (\theta_p \, z)^2$ represents the variance due to the pointing error, $\theta_p$ denotes the pointing jitter, and $\sigma_{\text{TB}}^2 \triangleq 0.1337\lambda^2 z^2w^{-1/3} r_c^{-5/3}$ represents the variance arising due to the beam wandering effect, with $ r_c = \left(0.423 \, k^2 C_n^2 \, z \right)^{-3/5}$ denoting the Fried parameter \cite{MIMOFSO_liu2020_9050208, pirandola2021satellite, Andrews2005}. Further, $C_n^2$ refers to the refractive index structure constant, which measures the level of turbulence following the Kolmogorov model. Generally, $C_n^2$ ranges from $10^{-14}\,\text{m}^{-2/3}$ (indicating moderate turbulence) to $10^{-17}\,\text{m}^{-2/3}$ (indicating weak turbulence) \cite{neel_IRS_assisted_nlosqkd_10289124}.
The displacement due to the misalignment factor typically follows a Weibull distribution \cite{weibull_4267802}, implying that the probability density function (p.d.f.) of $\tilde{r}$ is mathematically expressed as:
\beq
f_{\tilde{r}}(v) \triangleq \frac{v}{\sigma_{\tilde{r}}^2}
\exp \left(-\frac{v^2}{2\sigma_{\tilde{r}}^2}\right)
\, , \ v \geq 0 \, .
\label{eq6}
\eeq

Let the singular value decomposition (SVD) of the channel matrix generated using \eqref{eq1} be expressed as $\mathbf{H} = \mathbf{U} \mathbf{\Sigma} \mathbf{V}^\dagger$, where $\mathbf{U} \in \mathbb{C}^{N_R \times N_R}$ and $\mathbf{V} \in \mathbb{C}^{N_T \times N_T}$ are unitary matrices, with the diagonal matrix $\mathbf{\Sigma} \in \mathbb{R}^{N_R \times N_T}$ given as
\beq
\mathbf{\Sigma} \triangleq
\begin{bmatrix}
    \text{diag}\left(\sqrt{\beta_{1}},\ldots ,\sqrt{\beta_{r_H}}\right)&\boldsymbol{0}_{N_T \times\left(N_{T}-r_H\right)}\\
    \boldsymbol{0}_{\left(N_{R} -r_H\right) \times r_H}&\boldsymbol{0}_{\left(N_{R} -r_H\right) \times (N_{T}-r_H)}
\end{bmatrix} \, ,
\label{eq7}
\eeq 
where $r_H \leq \min\left( N_{T}, N_{R}\right)$ is the rank of $\mathbf{H}$ and the entries $\sqrt\beta_i$s, representing the transmittance of each $i$-th sub-channel ($i=1,\ldots,r_H$), are the non-zero singular values of $\mathbf{H}$. It is noted that the channel model in \eqref{eq1} incorporates the effect of Gaussian beam propagation, beam spreading, beam wandering, and the effect of the pointing error. However, apart from these effects, the transmittance of the FSO channel also depends on the atmospheric absorption, turbulence-induced fading, and detector efficiency. Thus, the effective transmissivity of each $i$-th sub-channel can be modified as follows:
\beq
T_i \triangleq \eta T_{a_i} T_{t_i} \beta_i,
\label{eq8}
\eeq
where $T_{a_i} \triangleq 10^{-\frac{\delta}{10} z}$ represents the attenuation due to atmospheric absorption, $\delta$ (in dB/m) is the attenuation coefficient, and $\eta$ denotes the detection efficiency at the receiver end. Additionally, $T_{t_i}$ captures the random fading caused by atmospheric turbulence for each $i$-th transmission path. Experimental results have shown that for long-distance quantum links, this random turbulence-induced fading follows a lognormal distribution \cite{Capraro_turbulance_2012}, implying that its p.d.f. is given by:
\beqarr
f_{T_{t_i}}(t_i) \triangleq \frac{1}{t_i \sqrt{2\pi \sigma^2}}
\exp \left( -\frac{\left( \ln(t_i) + \tfrac{\sigma^2}{2} \right)^2}{2\sigma^2} \right),
\label{eq9}
\eeqarr
where the parameter $\sigma^2$ represents the log-irradiance variance, characterizing the strength of turbulence, and is given by"
\beqarr
&& \!\!\!\!\!\!\!\!\sigma^2 \triangleq \exp \left(\frac{0.49 \chi^2}{\big( 1 + 0.18 d^2 + 0.56 \chi^{12/5} \big)^{7/6}} \right. \nn\\
&&\left. \qquad \quad + \frac{0.51 \chi^2}{\big( 1 + 0.9 d^2 + 0.62 d^2 \chi^{12/5} \big)^{5/6}}\right) - 1 ,
\label{eq10}
\eeqarr
where $\chi^2 \triangleq 1.23 C_n^2 k^{7/6} z^{11/6}$ and $d \triangleq a_r \sqrt{k/z}$.
\subsection{Additive Noise Model}
Apart from the atmospheric effect, the transmission of secret keys encoded using CV-QKD results in an additive noise occurring at the receiver end. For each $i$-th sub-channel, the noise, denoted by $n_{\text{GM}_i}$, follows the distribution of hybrid quantum noise, whose p.d.f. is derived from a Poisson–Gaussian mixture model \cite{mouli_jan2024_HQN_Goussian_channel_10403910}, \cite{mouli_hybridnoise_Jan2025}. This implies that we can express $n_{\text{GM}_i} = n_{\text{p},i} + n_{\text{g},i}$, where $n_{\text{p},i}$ represents the quantum Poisson noise and $n_{\text{g},i}$ indicates the classical additive white Gaussian noise (AWGN). The Poisson component models photon-counting noise due to background radiation and shot noise, while the Gaussian component captures thermal noise, electronic noise, and excess noise introduced by imperfect detectors. Such a Poisson–Gaussian noise model is standard in low-light optical and quantum receivers \cite{7553489}. Thus, the probability mass function (p.m.f.) of $n_{\text{p},i}$ is given by
\beq
f_{\text{p},i}(k,\lambda_0) \triangleq \frac{e^{-\lambda_0}\, \lambda_0^k}{k\, !} \, , \lambda_0\geq 0,\,k \in \{0,1,\ldots, \infty\}.
\label{eq11}
\eeq
Furthermore, $n_{\text{g},i}$ follows a Gaussian distribution with mean $\mu_{\text{g}}$ and variance $\sigma_{\text{g}}^2$, implying that $n_{\text{g},i} \sim {\mathcal{N}} \left(\mu_{\text{g}} , \sigma_{\text{g}}^2 \right)$ $\forall i=1,\ldots,r_H$.
Thus, the distribution of the hybrid quantum noise $n_{\text{GM}_i}$ can be computed as the convolution of the distributions of $n_{\text{p},i}$ and $n_{\text{g},i}$, yielding the expression: 
\beqarr
&& \hspace{-1.0cm}
f_{n_{\text{GM}_i}}(n) 
\triangleq \sum_{k=0}^{\infty} \frac{e^{-\lambda_0} \lambda_0^k}
{k! \sqrt{2\pi \sigma_{\text{g}}^2}} 
\exp \left(-\frac{\left(n-k-\mu_{\text{g}} \right)^2} 
{2\sigma_{\text{g}}^2} \right).
\label{eq12}
\eeqarr
\section{One- and Two-Way MIMO CV-QKD Protocols}
This section describes the considered one- and two-way protocols with their corresponding secret key transmission and reception mechanisms. In both protocols, the secret keys are transmitted by using the CV-QKD scheme, and Eve attempts to steal them applying a collective Gaussian attack. This attack is known to be the most comprehensive and effective attack strategy among the class of Gaussian protocols~\cite{navascues2006optimality}.
\subsection{One-Way Protocol}
\begin{figure}[!t]
    \centering
    \includegraphics[width=8.5cm,height=5cm]{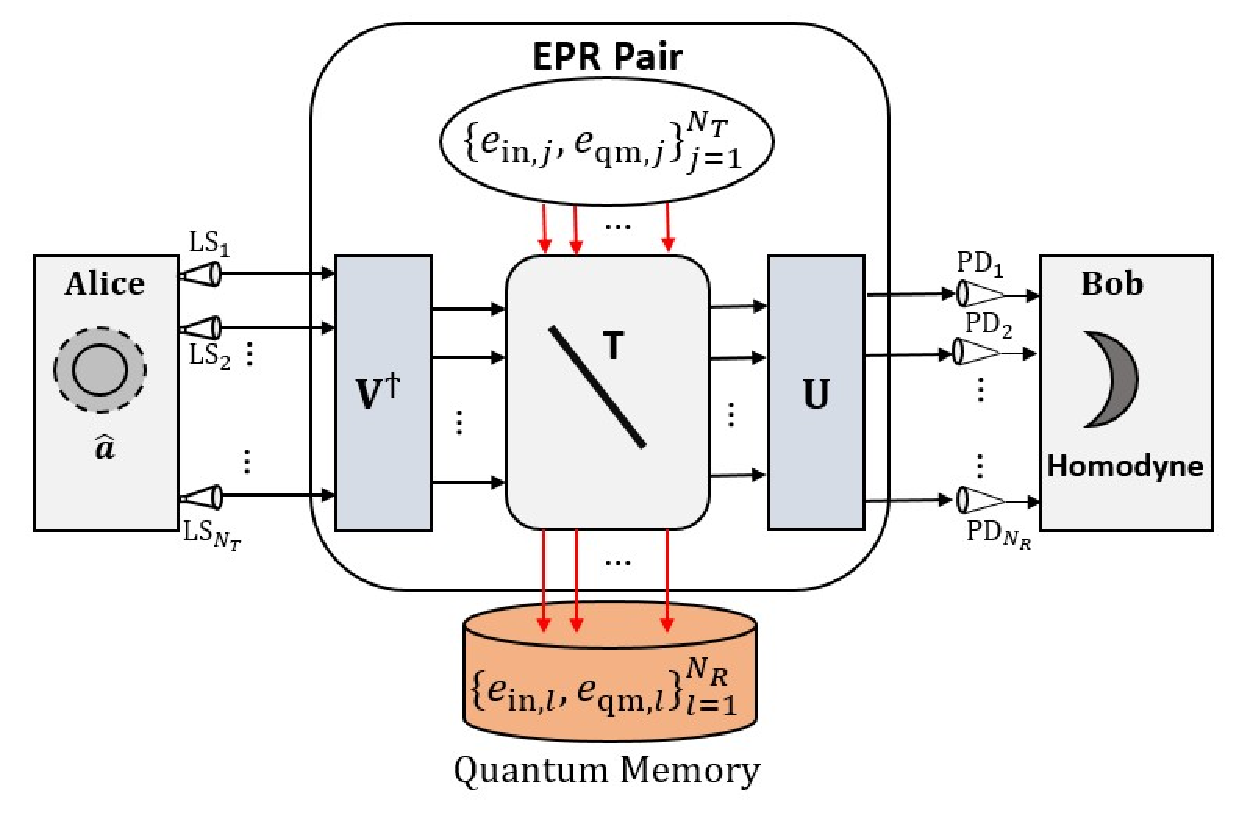}
    \caption{The MIMO FSO CV-QKD system model employing the one-way protocol.}
    \label{f2}
\end{figure}
As depicted in Fig.~\ref{f2}, in the one-way MIMO FSO communication system, Alice employs a Gaussian-modulated CV-QKD scheme to securely send secret keys to the legitimate receiver Bob. This transmission occurs over $\mathbf{H}$ that is vulnerable to eavesdropping, and thus, can be fully accessed by Eve. In this protocol, Alice prepares and transmits coherent states as $a_{j} \triangleq Q_{A, j} + \jmath P_{A, j}$ $\forall j = 1,\ldots, N_T$ from her $N_T$ LSs. Here, $Q_A$ and $P_A$ represent the position and momentum quadratures, respectively, at Alice's end. These quadratures are sampled independently from a zero-mean Gaussian distribution with a variance of $V_s$, implying that $Q_A, P_A \sim \mathcal{N}(0, V_s)$, with $\undb{E}[Q_A^2] = \undb{E}[P_A^2] = V_s$. Following this, Alice encodes her signals using the precoding matrix $\mathbf{V}$, while Bob employs the combining operation $\mathbf{U}^\dagger$ at their receiver to recover the secret keys. Further, the eavesdropper, Eve, executes a collective Gaussian attack for which a commonly utilized model is the entangling cloner, in which Eve prepares an Einstein–Podolsky–Rosen (EPR) state, denoted as $\rho_{e_{\text{in}} e_{\text{qm}}}$, as her quantum ancilla. This state represents a two-mode, zero-mean entangled Gaussian state, completely characterized by its covariance matrix, which is expressed as follows:
\beq
\mathbf{\Xi}_{e_{\text{in}} e_{\text{qm}}} \triangleq
\begin{bmatrix}
    \nu \mathbf{I_2} & \sqrt{\nu^2 - 1} \, \mathbf{Z} \\
    \sqrt{\nu^2 - 1} \, \mathbf{Z} & \nu \mathbf{I_2}
\end{bmatrix},
\label{eq13}
\eeq
where $\nu \geq 1$ is the variance of each mode (namely $e_{\text{in}}$ and $e_{\text{qm}}$) in the EPR state.

To compromise the transmission of the secret keys, Eve utilizes the collective Gaussian attack by coupling one mode from her EPR pair, referred to as $\mathbf{\hat{e}}_{\text{in}} \triangleq \left[ \hat{e}_\text{{in,1}}, \ldots, \hat{e}_{\text{in}, N_{T}} \right]^T$, with the input coherent state vector $\mathbf{\hat{a}} \triangleq \left[ \hat{a}_1, \ldots, \hat{a}_{N_{T}} \right]^T$. This is accomplished through a series of beam splitters characterized by the transmittance parameters $\{T_i\}_{i=1}^{r_H}$, which interact with the signals transmitted by Alice post using the precoding matrix $\mathbf{V}$. These beam splitters yield two distinct outputs for each input mode: \textit{i} the {\em legitimate channel output} which undergoes a unitary transformation $\mathbf{U}^{\dagger}$ at Bob's receiver end; and \textit{ii}) the {\em eavesdropper's output}, denoted by $\mathbf{e}_{\text{o}}$, which is stored in Eve's quantum memory along with the retained EPR mode $\mathbf{e}_{\text{q}}$. Eve delays her collective measurement until after Alice and Bob have concluded their classical communication phase, thereby optimizing her information gain. Thus, the effective mode transformation at Bob’s receiver can be represented as:
\beq
\mathbf{\hat{b}} \triangleq \mathbf{U}^{\dagger} \mathbf{H} \mathbf{V} \, \mathbf{\hat{a}} + \mathbf{U}^{\dagger} \mathbf{U} \mathbf{S} \, \mathbf{\hat{e}}_{\text{in}} +\mathbf{U}^{\dagger} \mathbf{\hat{n}}_{\text{GM}},
\label{eq14}
\eeq
where $\mathbf{\hat{b}} \triangleq \left[ \hat{b}_1, \ldots, \hat{b}_{N_{R}} \right]^T$ is the received quantum state vector, $\mathbf{\hat{n}}_{\text{GM}} \triangleq \left[ \hat{n}_{\text{GM}_1}, \ldots, \hat{n}_{\text{GM}_{N_R}} \right]^T$ is the additive hybrid noise vector, and $\mathbf{S} \in \mathbb{R}^{N \times N}$ is given by ($N \triangleq \min\{N_T, N_R\}$):
\beq
\textbf{S} \triangleq
\text{diag}\left(\sqrt{1-T_{1}}, \ldots, \sqrt{1-T_{r_H}}, \mathbf{1}_{\left(N-{r_H}\right)} \right).
\label{eq15}
\eeq

Applying the SVD of $\mathbf{H}$ in \eqref{eq15} and the overall transmissivity of the FSO channels in \eqref{eq8}, the input-output relation for each $i$-th link between Alice and Bob is obtained as:
\beq
\hat{b}_{i} =\sqrt{T_i}\, \hat{a}_{i} + \sqrt{1-T_i}\, \hat{e}_{{\text{in}_i}} +\hat{n}_{\text{GM}_i} \,
\forall i=1,\ldots, r_H.
\label{eq16}
\eeq
Additionally, Eve's output for each $i$-th link is given by:
\beq
\hat{e}_{{\text{o}_i}} = -\sqrt{1-T_i} \, \hat{a}_i +\sqrt{T_i} \,  \hat{e}_{{\text{in}_i}} \,
\forall i=1,\ldots, r_H.
\label{eq17}
\eeq
\subsection{Two-Way Protocol}
\begin{figure}
    \centering
    \includegraphics[width=8.5cm,height=6cm]{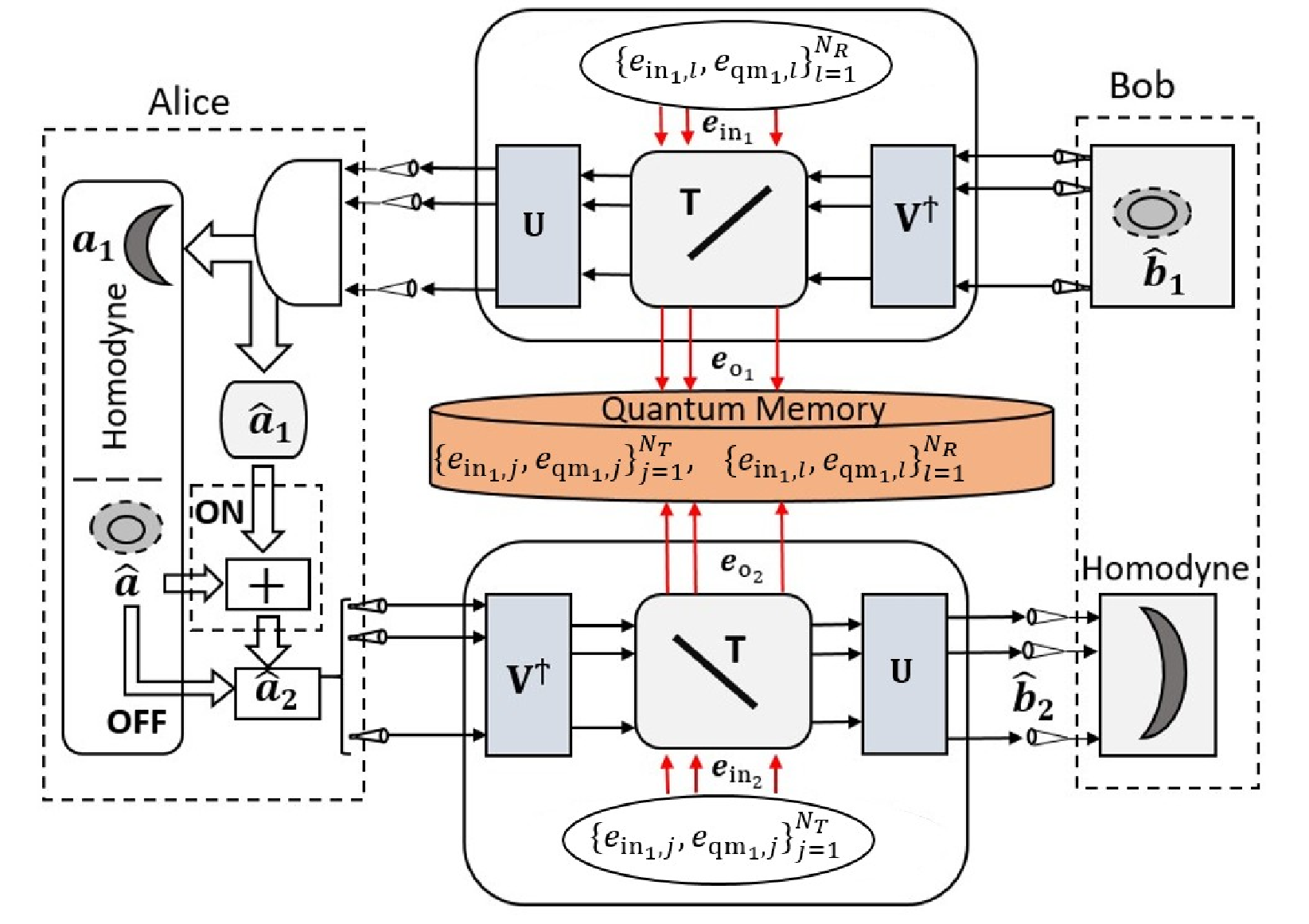}
    \caption{The MIMO FSO CV-QKD system model with the two-way protocol.}
    \label{f3}
\end{figure}
The system setup for the one-way protocol can be modified to implement a two-way protocol for quantum communications, as shown in Fig.~\ref{f3}. The two-way quantum communication protocol improves secure key distribution by utilizing bidirectional quantum channels between the legitimate parties. To implement this protocol, Alice and Bob are both equipped with LSs and PDs and, unlike the one-way protocol, Bob begins the process by transmitting $N_R$ Gaussian-modulated thermal coherent states $b_{1,l} \triangleq X_{b_1,l} + \jmath P_{b_1,l}$ $\forall l = 1,\ldots, N_R$ to Alice by utilizing a precoder matrix $\mathbf{V}$. Here, $X_{b_1,l}$ and $P_{b_1,l}$ are the position and momentum quadratures, respectively, following respectively the zero-mean Gaussian distributions $X_{b_1,l}, P_{b_1,l} \sim {\mathcal{N}} \left( 0, V_s \right)$. The signal mode travels via the quantum channel and reaches Alice, where she employs a combiner matrix $\mathbf{U}^\dagger$ to the receiving signal mode, which receives the noisy signal modes, denoted as  $a_{1,i}$s. During this transmission, Eve uses the Gaussian collective attack to steal the secret keys by following a process similar to the one described in the one-way protocol. Thus, the received signal modes at Alice are given as follows:
\beq
\hat{a}_{1,i} = \sqrt{T_i}\, \hat{b}_{1,i}
+ \sqrt{1-T_i}\, \hat{e}_{\text{in}_{1},i} 
+\hat{n}_{\text{GM}_i}^{(1)} \,
\forall i=1,\ldots, r_H, 
\label{eq18}
\eeq
where $\hat{n}_{\text{GM}_i}^{(1)}$ is the hybrid noise at Alice's end. On the other hand, Eve's received signal mode is given as
\beq
\hat{e}_{\text{o}_{1,i}} = -\sqrt{1-T_i}\, \hat{b}_{1,i} + \sqrt{T_i}\, \hat{e}_{\text{in}_{1,i}} \, \forall i=1,\ldots,r_H.
\label{eq19}
\eeq

Following this phase, Alice can randomly choose between two configurations, namely the \textbf{ON} and \textbf{OFF} configurations for secret key transmission to Bob. In the \textbf{ON} configuration, Alice creates her Gaussian states as $a_i = Q_{a, i} + \jmath P_{a,i}$, where $Q_{a, i}$ and $P_{a, i}$ are random values drawn from a Gaussian distribution with zero mean and a variance of $V_s$. She then transmits the updated mode as $\hat{a}_{2,i}=\hat{a}_{1,i} + \hat{a}_{i}$ $\forall i = 1\ldots,r_H$, back to Bob's end. On the contrary, in the \textbf{OFF} configuration, Alice performs a homodyne measurement on the incoming quantum states from Bob. Following this measurement, she prepares new coherent states represented by $\hat{a}_{2, i}=\hat{a}_{i}$ $\forall i = 1, \ldots, r_H$ and transmits them to Bob. At the receiving end, Bob applies his detection technique to the incoming mode.

At the conclusion of the two-way quantum communication phase, Alice and Bob perform classical post-processing over an authenticated public channel. In this phase, Alice announces the configuration, i.e., either \textbf{ON} or \textbf{OFF}, that was selected in each round of the secret key exchange. Furthermore, both parties disclose which quadratures were measured by their respective detection technique. This public exchange enables them to identify the correlated measurement outcomes. In the \textbf{OFF} configuration, Alice and Bob acquire two independent sets of correlated variables, denoted as $\{a_{1, i}, b_{1, i}\}_{i=1}^{r_H}$ and $\{a_{2, i}, b_{2, i}\}_{i=1}^{r_H}$. In contrast, in the \textbf{ON} configuration, they obtain a single correlated pair $\{a_{i}, b_{i}\}_{i=1}^{r_H}$, where Bob's variable $b_i$ is constructed from the linear combination of $b_{1, i}$ and $b_{2, i}$ is obtained during the Bob-to-Alice and Alice-to-Bob signal transmissions, respectively.


The \textbf{ON} configuration has been shown to remain effective even in regimes where one-way communication protocols fail \cite{Pirandola_2008,twoway_weedbrook_2013}. For this reason, we assume Alice selects this configuration and, in the second phase of transmission, she applies the precoding matrix $\mathbf{V}$ to encode her quantum states, while Bob employs the combining matrix $\mathbf{U}^\dagger$ for signal recovery. As in the previous phase, the communication channel remains vulnerable to eavesdropping, and the signal modes received by Bob are modeled accordingly to reflect this potential interception, which is given as follows:
\beq
\hat{b}_{2,i} = \sqrt{T_i} \hat{a}_{2,i} + \sqrt{1-T_i}\hat{e}_{\text{in}_{2,i}} + \hat{n}_{\text{GM}_i}^{(2)} \, \forall i=1,\ldots,r_H,
\label{eq20}
\eeq
where $\hat{n}_{\text{GM}_i}^{(2)}$ is the hybrid noise at Bob's end in the $i$-th sub-channel. Moreover, Eve's received signal is obtained as:
\beq
\hat{e}_{\text{o}_{2,i}} = -\sqrt{1-T_i} \hat{a}_{2,i} + \sqrt{T_i}\,\hat{e}_{\text{in}_{2,i}} \, \forall i=1,\ldots,r_H.
\label{eq21}
\eeq
\section{Analysis of the Mutual Information}
The performance of the MIMO FSO CV-QKD systems under consideration is henceforth characterized with respect to their SKRs. The computation of the SKR involves the computation of the classical mutual information between Alice and Bob, which is analytically assessed for both the one- and two-way protocols in this section.
\subsection{One-Way Protocol}
At the receiver end in the CV-QKD system when employing the one-way protocol, Bob employs homodyne detection on each  received signal mode $\hat{b}_i$ (recall $i= 1 ,\ldots, r_H$), given in \eqref{eq16}, resulting in the following output with $a_i \sim \mathcal{N}(0, V_s)$: 
\beq
b_{i}^{\text{1-way}} = \sqrt{T_i}\, a_{i} 
+ n_{\text{GM}_i}^{\text{1-way}},
\label{eq22}
\eeq
where the term $n_{\text{GM}_i}^{\text{1-way}} = \sqrt{T}\,a_0+\sqrt{1-T_i}\, e_{{\text{in}}_i} + n_{\text{GM}_i}$ is the effective additive noise, with $a_0$ representing the vacuum thermal noise at Alice following a zero-mean Gaussian distribution with a variance of $V_0$ (i.e., $a_0 \sim {\mathcal{N}} \left(0, V_0 \right)$), and $e_{{\text{in}}_i}\sim \mathcal{N}(0, \nu)$ being the eavesdropping signal. Using these statistics and the statistics of the hybrid quantum noise given in \eqref{eq12}, the p.d.f. of $n_{\text{GM}_i}^{\text{1-way}}$ can be obtained as follows:
\beq
f_{n_{\text{GM}_i}^{\text{1-way}}}(n) =
\sum_{k=0}^{\infty} \frac{e^{-\lambda_0} \lambda_0^k}
{k! \sqrt{2\pi \sigma_{n_i,\text{1-way}}^2}} 
\exp\left(-\frac{\left(n-k-\mu_{\text{g}}\right)^2} {2\sigma_{n_i,\text{1-way}}^2}  \right), 
\label{eq23}
\eeq
where $\sigma_{n_i,\text{1-way}}^2 = T_iV_0+\left(1-T_i\right)\nu+ \sigma_{\text{g}}^2$. Thus, from \eqref{eq22} and \eqref{eq23}, the p.d.f. of the received signal at Bob's end is given by
\beq
f_{b_{i}^{\text{1-way}}}(b) = \sum_{k=0}^{\infty} \frac{e^{-\lambda_0} \lambda_0^k}{k!\sqrt{2\pi \sigma_{b_i,\text{1-way}}^2}} 
\exp\left(- \frac{\left(b - k - \mu_{\text{g}}\right)^2} {2\sigma_{b_i,\text{1-way}}^2}\right),
\label{eq24}
\eeq
where $\sigma_{b_i,\text{1-way}}^2 = T_i V_a+\left(1-T_i\right)\nu+ \sigma_{\text{g}}^2$ and $V_a=V_s+V_0$.

Following the latter derivations, the classical mutual information between the transmitted signal $a_i$ and the received signal $b_{i}^{\text{1-way}}$, denoted by $I\left(a_i; b_{i}^{\text{1-way}} \right)$,
can be computed as:
\begin{align}
& \! \! \! \! \! \! \! \!
I\left(a_i; b_{i}^{\text{1-way}} \right)
= h\left(a_i\right) + h\left( b_{i}^{\text{1-way}} \right)
- h \left(a_i, b_{i}^{\text{1-way}} \right) \nn \\
&= h\left(a_i\right) + h\left(b_{i}^{\text{1-way}} \right)
- \left(h\left(b_{i}^{\text{1-way}} \big| a_i\right)
+ h\left(a_i\right) \right)\nn \\
&= h\left(b_{i}^{\text{1-way}} \right)
- h\left(\sqrt{T_i}\, a_{i} 
+ n_{\text{GM}_i}^{\text{1-way}} \big| a_i\right)\nn \\
&= h\left(b_{i}^{\text{1-way}} \right)
- h \left(n_{\text{GM}_i}^{\text{1-way}} \right),
\label{eq25}
\end{align}
where $h\left(b_{i}^{\text{1-way}}\right)$ and $h\left(n_{\text{GM}_i}^{\text{1-way}} \right)$ denote the differential entropies of the received signal $b_{i}^{\text{1-way}}$ and the effective additive noise $n_{\text{GM}_i}^{\text{1-way}}$ for each $i$-th sub-channel, respectively. The quantities are analytically derived as follows:
\begin{align}
&h \left( b_{i}^{\text{1-way}} \right)
= - \int_{-\infty}^{\infty}
f_{b_{i}^{\text{1-way}}}(b)
\log\left( f_{b_{i}^{\text{1-way}}}(b) \right)
\, \text{d}b \nn \\
& = - \int_{-\infty}^{\infty}
\left(\sum_{k=0}^{\infty} \frac{e^{-\lambda_0} \lambda_0^k}
{k! \sqrt{2\pi \sigma_{b_i,\text{1-way}}^2}} 
e^{- \frac{\left(b - k - \mu_{\text{g}}\right)^2} {2\sigma_{b_i,\text{1-way}}^2}}\right) \nn\\
& \hspace{0.37cm} \times \log \! \left( \sum_{k=0}^{\infty} \frac{e^{-\lambda_0} \lambda_0^k}
{k! \sqrt{2\pi \sigma_{b_i,\text{1-way}}^2}} 
e^{- \frac{\left(b - k - \mu_{\text{g}}\right)^2} {2\sigma_{b_i,\text{1-way}}^2}} \right) \text{d} b,
\label{eq26}
\end{align}
\begin{align}
&h\left(n_{\text{GM}_i}^{\text{1-way}} \right)
= - \int_{-\infty}^{\infty}
f_{n_{\text{GM}_i}^{\text{1-way}}}(n)
\log\left( f_{n_{\text{GM}_i}^{\text{1-way}}}(n) \right)
\, \text{d}n \nn \\
& = - \int_{-\infty}^{\infty}
\left( \sum_{k=0}^{\infty} \frac{e^{-\lambda_0} \lambda_0^k}
{k! \sqrt{2\pi \sigma_{n_i,\text{1-way}}^2}} 
e^{\left(-\frac{\left(n-k-\mu_{\text{g}}\right)^2} {2\sigma_{n_i,\text{1-way}}^2}  \right)} \right) \nn\\
& \hspace{0.37cm} \times \log \! \left( \sum_{k=0}^{\infty} \frac{e^{-\lambda_0} \lambda_0^k}
{k! \sqrt{2\pi \sigma_{n_i,\text{1-way}}^2}} 
e^{\left(-\frac{\left(n-k-\mu_{\text{g}}\right)^2} {2\sigma_{n_i,\text{1-way}}^2}  \right)} \right) \text{d} n.
\label{eq27}
\end{align}

It can be observed from \eqref{eq26} and \eqref{eq27} that the differential entropies involve infinite summations of weighted exponential functions; thus, their numerical evaluation can be mathematically intractable. 
To this end, we next present tight bounds on these expressions, leading to a bound on the system's mutual information. 
We particularly perform the definitions:
\begin{subequations}\label{eq28}
\begin{equation}\label{eq28a}
h_{\text{L}} \left(n_{\text{GM}_i}^{\text{1-way}}\right)
\leq h\left(n_{\text{GM}_i}^{\text{1-way}}\right)
\leq h_{\text{U}} \left(n_{\text{GM}_i}^{\text{1-way}}\right),
\end{equation}
\begin{equation}\label{eq28b}
h_{\text{L}} \left(b_{i}^{\text{1-way}}\right)
\leq h\left(b_{i}^{\text{1-way}}\right)
\leq h_{\text{U}} \left(b_{i}^{\text{1-way}}\right).
\end{equation}
\end{subequations}
A bound on $I\left(a_i; b_{i}^{\text{1-way}} \right)$ can be obtained as follows:
\begin{align}
& h_{\text{L}} \left(n_{\text{GM}_i}^{\text{1-way}}\right) 
\leq h \left(n_{\text{GM}_i}^{\text{1-way}}\right)
\leq h_{\text{U}} \left(n_{\text{GM}_i}^{\text{1-way}}\right) 
\nn \\
\implies
& - h_{\text{L}} \left(n_{\text{GM}_i}^{\text{1-way}}\right) 
\geq - h \left(n_{\text{GM}_i}^{\text{1-way}}\right)
\geq - h_{\text{U}} \left(n_{\text{GM}_i}^{\text{1-way}}\right) 
\nn \\
\implies
& h \left( b_{i}^{\text{1-way}} \right)
- h_{\text{L}} \left(n_{\text{GM}_i}^{\text{1-way}}\right)
\geq  h \left( b_{i}^{\text{1-way}} \right)
- h \left(n_{\text{GM}_i}^{\text{1-way}}\right) \nn \\ 
& \qquad \qquad \qquad \qquad \qquad \,
\geq h \left( b_{i}^{\text{1-way}} \right)
- h_{\text{U}} \left(n_{\text{GM}_i}^{\text{1-way}}\right) \nn\\
\implies
& h_{\text{U}} \left( b_{i}^{\text{1-way}} \right)
- h_{\text{L}} \left(n_{\text{GM}_i}^{\text{1-way}}\right)
\geq h \left( b_{i}^{\text{1-way}} \right)
- h_{\text{L}} \left(n_{\text{GM}_i}^{\text{1-way}}\right) \nn\\
& \qquad \qquad \qquad \qquad \qquad \ \
\geq h \left(b_{i}^{\text{1-way}}\right)
- h \left(n_{\text{GM}_i}^{\text{1-way}}\right) \nn\\
& \qquad \qquad \qquad \qquad \qquad \ \
= I\left(a_i; b_{i}^{\text{1-way}} \right).
\label{eq29}
\end{align}
Furthermore, the expressions for $h_{\text{U}} \left( b_{i}^{\text{1-way}} \right)$ and $h_{\text{L}} \left(n_{\text{GM}_i}^{\text{1-way}}\right)$ can be respectively derived using \cite[Theorems 2,3]{4648062} as:
\begin{align}
\! \! h_{\text{U}} \left( b_{i}^{\text{1-way}} \right)
= \sum_{k=0}^{\infty} \frac{e^{-\lambda_0} \lambda_0^k}{k!}
& \left( - \log_2 \left(\frac{e^{-\lambda_0} \lambda_0^k}{k!} \right) \right. \nn \\
& \quad \left. + \frac{1}{2} \log_2 \left(2\pi e \sigma_{b_i,\text{1-way}}^2 \right) \right),
\label{eq30}
\end{align}
\begin{align}
h_{\text{L}} \left(n_{\text{GM}_i}^{\text{1-way}}\right)
= - & \sum_{k=0}^{\infty}
\left[\frac{e^{-\lambda_0} \lambda_0^k}{k!}
\log_2 \left(\sum_{\ell=0}^{\infty} \frac{e^{-\lambda_0} \lambda_0^{\ell}}
{\ell! \sqrt{4 \pi \sigma_{n_i,\text{1-way}}^2}} 
\right. \right.\nn \\
& \left. \left. \qquad \qquad
\times \exp\left(-\frac{\left(k -\ell \right)^2}
{4\sigma_{n_i,\text{1-way}}^2} \right) \right) \right].
\label{eq31}
\end{align}
Substituting \eqref{eq30} and \eqref{eq31} in \eqref{eq29}, and performing some algebraic simplifications, the following upper bound on the mutual information for the MIMO FSO CV-QKD system employing the one-way protocol is obtained:
\begin{align}
&I_{\text{B}} \left(a_i; b_{i}^{\text{1-way}} \right) = \sum_{k=0}^{\infty}
\frac{e^{-\lambda_0} \lambda_0^k}{k!}
\left( -\log_2\left(\frac{e^{-\lambda_0} \lambda_0^k}{k!}\right)\right. \nn\\
& + \frac{1}{2} \log_2 \left(2 \pi e \left(T_i V_a
+ \left(1-T_i \right) \nu
+ \sigma_{\text{g}}^2 \right) \right) \nn \\
& + \log_2 \left(\sum_{\ell=0}^{\infty}
\frac{e^{-\lambda_0} \lambda_0^{\ell}}
{\ell! 2\sqrt{\pi\left(T_iV_0+\left(1-T_i\right)\nu
+ \sigma_{\text{g}}^2\right)}} \right. \nn \\
& \quad \left. \left.
\times \exp \left(- \frac{\left(k -\ell \right)^2}
{4\left(T_iV_0+\left(1-T_i\right) \nu
+ \sigma_{\text{g}}^2\right)}  \right) \right) \right).
\label{eq32}
\end{align}
\subsection{Two-Way Protocol}
For the case where the considered MIMO FSO CV-QKD system employs the $\textbf{ON}$ configuration with the two-way protocol, the received signal mode at Bob is given in \eqref{eq20}. Then, this node employs the homodyne detection to each received signal mode $\hat{b}_{2, i}$, which is then followed by the post processing of subtracting the input modulation $b_{1, i}$ to generate the effective variable $b_i$~\cite{twoway_weedbrook_2013}, as follows $\forall i = 1, \ldots, r_H$: 
\beqarr
b_i \! \! \! \! &=& \! \! \! \! b_{2,i} - T_i b_{1,i} \nn \\
&=& \! \! \! \! \sqrt{T_i} a_i + T_i b_{0} + \sqrt{1 - T_i} \left( \sqrt{T_i} \, e_{\text{in}_{1,i}} + e_{\text{in}_{2,i}} \right) \nn \\
&& \! \! \! + \sqrt{T_i}\, n_{\text{GM}_i}^{(1)}+ n_{\text{GM}_i}^{(2)},
\label{eq33}
\eeqarr 
where $n_{\text{GM}_i}^{(1)}$ and $n_{\text{GM}_i}^{(2)}$ are considered to be statistically independent of each other. This expression can be re-written as:
\beq
b_i = \sqrt{T_i}\,a_{i} + n_{\text{GM}_i}^{\text{2-way}},
\label{eq34}
\eeq
where $n_{\text{GM}_i}^{\text{2-way}} = \sqrt{\left(1 - T_i\right)} \left(\sqrt{T_i} \, e_{\text{in}_{1, i}} + e_{\text{in}_{2, i}} \right) + \sqrt{T_i}a_0 + T_i b_{0} + \sqrt{T_i}\, n_{\text{GM}_i}^{(1)}+ n_{\text{GM}_i}^{(2)}$ represents the effective hybrid quantum noise arising from Alice's and Bob's vacuum thermal noise, the eavesdropping signals, and hybrid quantum noise in the two-way framework, with $b_0 \sim {\mathcal{N}} \left(0, V_0 \right)$. From the statistics of all the involved terms, the p.d.f. of $n_{\text{GM}_i}^{\text{2-way}}$ can be obtained as follows:
\begin{align}
f_{n_{\text{GM}_i}^{\text{2-way}}}(n)
= & \sum_{k=0}^{\infty} \frac{e^{-\lambda_0} \lambda_0^k}
{k! \sqrt{2\pi\sigma_{n_i,\text{2-way}}^2}} \nn \\
& \times
\exp \left( -\frac{\left(n-k-\left(1+\sqrt{T_i}\right)\mu_{\text{g}} \right)^2} 
{2\sigma_{n_i,\text{2-way}}^2} \right),
\label{eq35}
\end{align}
where $\sigma_{n_i,\text{2-way}}^2 = \left(T_i+T_i^2\right) V_0 +\left(1-T_i^2\right)\nu+\left(1+T_i^2\right)\sigma_{\text{g}}^2$. 
Moreover, the p.d.f. of the received signal at Bob's end can be computed using \eqref{eq34} and \eqref{eq25} as:
\begin{align}
f_{b_i^{\text{2-way}}}(b)
= & \sum_{k=0}^{\infty} \frac{e^{-\lambda_0} \lambda_0^k}
{k! \sqrt{2\pi \sigma_{b_i,\text{2-way}}^2}} \nn \\
& \times 
\exp \left(- \frac{\left(b-k- \left(1+\sqrt{T_i}\right)\mu_{\text{g}}\right)^2} 
{2\sigma_{b_i,\text{2-way}}^2} \right),
\label{eq36}
\end{align}
where $\sigma_{b_i,\text{2-way}}^2 = T_i V_a+T_i^2 V_0 + \left(1-T_i^2\right) \nu + \left(1+T_i^2\right)\sigma_{\text{g}}^2$. 

It is noted that the expressions~\eqref{eq35} and~\eqref{eq36} are similar to those obtained for the one-way protocol in Section III.A. Thus, we can similarly compute the upper bound on the differential entropy of $b_i^{\text{2-way}}$, denoted by $h_{\text{U}} \left(b_i^{\text{2-way}}\right)$, and the lower bound on the differential entropy of $n_{\text{GM}_i}^{\text{2-way}}$, denoted by $ h_{\text{L}} \left(n_{\text{GM}_i}^{\text{2-way}}\right)$:
\begin{align}
\! \! h_{\text{U}} \left( b_{i}^{\text{2-way}} \right)
= \sum_{k=0}^{\infty} \frac{e^{-\lambda_0} \lambda_0^k}{k!}
& \left( - \log_2 \left(\frac{e^{-\lambda_0} \lambda_0^k}{k!} \right) \right. \nn \\
& \quad \left. + \frac{1}{2} \log_2 \left(2\pi e \sigma_{b_i,\text{2-way}}^2 \right) \right) ,
\label{eq37}
\end{align}
\begin{align}
h_{\text{L}} \left(n_{\text{GM}_i}^{\text{2-way}}\right)
= - & \sum_{k=0}^{\infty}
\left[\frac{e^{-\lambda_0} \lambda_0^k}{k!}
\log_2 \left(\sum_{\ell=0}^{\infty} \frac{e^{-\lambda_0} \lambda_0^{\ell}}
{\ell! \sqrt{4 \pi \sigma_{n_i,\text{2-way}}^2}} 
\right. \right.\nn \\
& \left. \left. \qquad \qquad
\times \exp\left(-\frac{\left(k -\ell \right)^2}
{4\sigma_{n_i,\text{2-way}}^2} \right) \right) \right] \, .
\label{eq38}
\end{align}
Using the latter expressions, the following upper bound on the mutual information between the transmitted signal $a_i$ and the received signal $b_i^{\text{2-way}}$ for the MIMO FSO CV-QKD system, for the case where the two-way protocol is used, is deduced:
\begin{align}
& I_{\text{B}} \left(a_i; b_{i}^{\text{2-way}} \right) \nn \\
& = \sum_{k=0}^{\infty} \frac{e^{-\lambda_0} \lambda_0^k}{k!}
\left(-\log_2\left(\frac{e^{-\lambda_0} \lambda_0^k}{k!}\right) 
 + \frac{1}{2} \log_2 \left(2\pi e \right. \right. \nn \\
& \left. \times \left(T_iV_a +T_i^2 V_0+\left(1-T_i^2\right) \nu + \left(1+T_i\right)\sigma_{\text{g}}^2\right)\right) \nn \\
&+\log_2 \left(\sum_{\ell=0}^{\infty} \frac{e^{-\lambda_0} \lambda_0^{\ell}}
{\ell! 2\sqrt{\pi\left(T_i^2 V_0 +\left(1-T_i^2\right)\nu+\left(1+T_i\right)\sigma_{\text{g}}^2\right)}}
\right. \nn\\
& \left. \left. \times
\exp \left( - \frac{\left(k - \ell \right)^2}
{4\left(T_i^2 V_0 +\left(1-T_i^2\right)\nu
+\left(1+T_i\right)\sigma_{\text{g}}^2\right)}
\right) \right) \right).
\label{eq39}
\end{align}
\section{SKR Analysis Under Hybrid Quantum Noise}
In response to Eve's Gaussian collective attack, Bob employs homodyne detection along with RR to correct any errors that may occur during transmission. Notably, the RR protocol ensures a positive SKR even in the low-transmissivity regime of $T_i$ $\in[0,1]$~\cite{weedbrook2010quantum} (high-loss). Following this setup, in this section, analytical SKR expression for both one- and two-way MIMO FSO CV-QKD systems are presented.
\subsection{One-Way MIMO FSO CV-QKD}
The effective SKR for each $i$-th ($i=1,\ldots, r_H$) channel between Alice and Bob for secret key exchange with the one-way protocol is expressed as follows:
\begin{align}
\text{SKR}^{\text{1-way}}_i &\triangleq 
\beta I \left(a_i ; b_i^{\text{1-way}} \right)
- \chi \left(e_i ; b_i^{\text{1-way}} \right) \nn \\
& \approx \beta I_{\text{B}} \left(a_i ; b_i^{\text{1-way}} \right)
- \chi \left(e_i;b_i^{\text{1-way}}\right),
\label{eq40}
\end{align}
where $\beta$ is the reconciliation efficiency factor and $I_{\text{B}} \left(a_i ; b_i^{\text{1-way}} \right)$ is given in \eqref{eq32}. In addition, $\chi \left(e_i ; b_i^{\text{1-way}} \right)$ denotes the Holevo information \cite{weedbrook2010quantum} between $e_i$ and $b_i^{\text{1-way}}$, which is given by:
\begin{align}
\chi\left(e_i; b_i^{\text{1-way}}\right) 
& \triangleq S \left(e_i\right) - S\left(e_i \big| b_i^{\text{1-way}}\right)
\nn \\
& = \sum_{q=1}^2 h_o \left(\lambda_{i_q}^{\text{1-way}}\right) 
- \sum_{q=3}^4 h_o\left(\lambda_{i_q}^{\text{1-way}}\right),
\label{eq41}
\end{align}
where $S(e_i)$ and $S\left(e_i \big| b_i^{\text{1-way}} \right)$ represent the Von Neumann entropy of Eve's ancillary state, which she stores in her quantum memory, and the conditional Von Neumann entropy of Eve's ancillary state given Bob's quadrature mode $b_i$. Both of these entropies depend on the symplectic eigenvalues, denoted as $\lambda_q$'s $(\geq 0)$, derived from the covariance matrix of the corresponding Gaussian states. Furthermore, the function $h_o(\cdot)$ used to calculate the Von Neumann entropy of a Gaussian quantum state is defined as follows~\cite{weedbrook2010quantum}:
\beq
h_o (\lambda_q ) \triangleq  \frac{\lambda_q+1}{2}
\log_2 \left( \frac{\lambda_q+1}{2} \right)
- \frac{\lambda_q-1}{2}
\log_2 \left( \frac{\lambda_q-1}{2} \right).
\label{eq42}
\eeq

Corresponding to each $i$-th link, Eve's ancillary state comprises the two modes $e_{\text{qm}, i}$ and $e_{\text{o},i}$. The related covariance matrix for these modes, denoted as $\mathbf{\Xi}_{E_i}^{\text{1-way}}$, is given as:
\beq
\mathbf{\Xi}_{E_i}^{\text{1-way}} \triangleq 
         \left[
         \begin{array}{cc}
             \Lambda_i\left(\nu,V_a\right) \, \mathbf{I}_2 & c_i \, \mathbf{Z} \\
             c_i \, \mathbf{Z}^T & \nu\, \mathbf{I}_2      
         \end{array}
         \right],
\label{eq43}
\eeq
where $\Lambda_i\left(\nu,V_a\right) \triangleq T_i \nu + \left(1-T_i\right)V_a$,  $V_a\triangleq V_s+V_0$, and $c_i \triangleq \sqrt{T_i (\nu^2 - 1)}$. The symplectic eigenvalues of this matrix are determined as the absolute values of the eigenvalues of $\jmath \mathbf{\Omega} \mathbf{\Xi}_{E_i}^{\text{1-way}}$, where $\mathbf{\Omega} \in \mathbb{R}^{2n \times 2n}$ is defined, for an $n$-mode system, as follows:
\beq
\mathbf{\Omega} \dn \bigoplus_{j=1}^n \boldsymbol{\omega} = \mathbf{I}_n \otimes \boldsymbol{\omega}\, , \quad \text{with } \boldsymbol{\omega} = 
\begin{bmatrix}
0 & 1 \\
-1 & 0
\end{bmatrix} .
\label{eq44}
\eeq
Additionally, the conditional covariance matrix $\mathbf{\Xi}_{E_{i}|b_i}^{\text{1-way}}$ is:
\beq
\mathbf{\Xi}_{E_{i}|b_i}^{\text{1-way}}
= \mathbf{\Xi}_{E_{i}}^{\text{1-way}}
-\frac{1}{V_{b_i}^{\text{1-way}}}
\mathbf{\Sigma}_{i}^{\text{1-way}} \mathbf{\Pi}
\left( \mathbf{\Sigma}_{i}^{\text{1-way}} \right)^T,
\label{eq45}
\eeq
where $V_{b_i}^{\text{1-way}} \triangleq  T_i V_a + \left(1-T_i\right)\nu +\lambda_0+\sigma_{\text{g}}^2 $ is the variance of the Bob's receiver signal $b_i$, $\mathbf{\Xi}_{E_{i}}^{\text{1-way}}$ is given in (\ref{eq43}), and $\mathbf{\Pi}$ and $\mathbf{\Sigma}_i^{\text{1-way}}$ are defined as:
\beq
\mathbf{\Pi} \dn
\begin{bmatrix}
    1 & 0 \\
    0 & 0
\end{bmatrix}, \quad
\mathbf{\Sigma}_i^{\text{1-way}} =
         \left[
         \begin{array}{c}
             \delta_{1,i} \, \mathbf{I}_2 \\
             \delta_{2,i} \, \mathbf{Z}
         \end{array}
         \right],
\label{eq46}
\eeq
where $\delta_{1,i} \triangleq  \sqrt{T_i\left(1-T_i\right)}\left(\nu-V_a\right)$ and $\delta_{2,i} \triangleq  \sqrt{\left(1-T_i\right) \left(\nu^2-1 \right)}$.
The symplectic eigenvalues can be derived from the latter results, and can be then used to obtain the analytical expression (\ref{eq47}) for the effective SKR of the one-way system (top of the next page).
\begin{figure*}[!t]
\beqarr
\text{SKR}_{\text{MIMO}}^{\text{1-way}}
= \sum_{i=1}^{r_H} \undb{E}_{T_i}
\left[ \text{SKR}_{i}^{\text{1-way}} \right]
\approx \sum_{i=1}^{r_H} \undb{E}_{T_i}
\left[ \beta \, I_{\text{B}} \left(a_i ; b_i^{\text{1-way}} \right)
- \sum_{q=1}^2 h_o \left( \lambda_{i_q}^{\text{1-way}} \right)
+ \sum_{q=3}^4 h_o \left( \lambda_{i_q}^{\text{1-way}} \right) \right]
\label{eq47}
\eeqarr
\noindent\rule{\textwidth}{.5pt}
\vspace{-0.8cm}
\end{figure*}
\subsection{Two-Way MIMO FSO CV-QKD}
For the two-way protocol, we consider that Alice chooses the \textbf{ON} configuration, in which she transmits a newly generated signal along with the originally received signal from Bob. When Bob receives this signal, he performs homodyne detection and RR. The SKR for each $i$-th LS and PD link between Alice and Bob is expressed as follows:
\beq
\text{SKR}^{\text{2-way}}_i
\approx \beta I_{\text{B}} \left(a_i ; b_i^{\text{2-way}} \right)
- \chi \left(e_i;b_i^{\text{2-way}}\right),
\label{eq48}
\eeq
where $I_{\text{B}} \left(a_i ; b_i^{\text{2-way}} \right)$ is computed in (\ref{eq39}) and the Holevo information is given as
\beqarr
\chi \left(e_i; b_i^{\text{2-way}} \right)
= \sum_{q=1}^4 h_o \left( \lambda_{i_q}^{\text{2-way}} \right)-\sum_{q=5}^8 h_o \left( \lambda_{i_q}^{\text{2-way}} \right).
\label{eq49}
\eeqarr
In this expression, $\lambda_{i_q}^{\text{2-way}}$'s represent the symplectic eigenvalues of the covariance matrix of Eve's overall quantum state, which comprises the modes $e_{\text{o}_1,i}$, $e_{\text{qm}_1,i}$, $e_{\text{o}_2,i}$, and $e_{\text{qm}_2,i}$. These eigenvalues can be determined in a manner analogous to that used in the one-way system. To this end, the covariance matrix of all quantum states available to Eve can be obtained as~\cite{twoway_weedbrook_2013}:
\beq
\mathbf{\Xi}_{E_i}^{\text{2-way}} \triangleq
         \left[
         \begin{array}{cccc}
             \sigma_{11_i} \mathbf{I}_2 & c_i \mathbf{Z} & \sigma_{21_i} \mathbf{I}_2 & \mathbf{0}_2 \\
             c_i \mathbf{Z}^T & \nu \mathbf{I}_2 & \bar{c}_i \mathbf{Z} & \mathbf{0}_2 \\
             \sigma_{21_i} \mathbf{I}_2 & \bar{c}_i \mathbf{Z}^T & \sigma_{22_i} \mathbf{I}_2 & c_i \mathbf{Z} \\
             \mathbf{0}_2 & \mathbf{0}_2 & c_i \mathbf{Z}^T & \nu \mathbf{I}_2
         \end{array}
         \right],
\label{eq50}
\eeq
where 
\begin{align}
\sigma_{11_i} \triangleq & \left(1 - T_i\right) V_a + T_i \nu , \nn\\
\sigma_{21_i} \triangleq & \sqrt{T_i}\left(1 - T_i\right)\left(V_a-\nu\right) , \nn \\
\bar{c}_i \triangleq & -\left(1- T_i\right)\sqrt{(\nu^2 - 1)} , \nn \\
\sigma_{22_i} \triangleq & \left(1 - T_i^2\right)V_a+\left(1-T_i+T_i^2\right)\nu
\nn \\
& + \left(1-T_i\right)\left(\lambda_0+\sigma_{\text{g}}^2\right).
\label{eq51}
\end{align}
Additionally, the conditional covariance matrix of Eve's quantum states for each $i$-th double connection between Alice and Bob can be calculated in a manner akin to (\ref{eq45}), as follows:
\beq
\mathbf{\Xi}_{E_{i}|b_i}^{\text{2-way}}
= \mathbf{\Xi}_{E_{i}}^{\text{2-way}}
-\frac{1}{V_{b_i}^{\text{2-way}}}
\mathbf{\Sigma}_{i}^{\text{2-way}} \mathbf{\Pi}
\left( \mathbf{\Sigma}_{i}^{\text{2-way}} \right)^T,
\label{eq52}
\eeq
where Bob's variance is given by $V_{b_i}^{\text{2-way}} = T_i V_a+T_i^2V_0+\left(1-T_i^2\right)\nu +\left(1+T_i\right)\lambda_0+\left(1+T_i\right)\sigma_{\text{g}}^2$ and the term $\mathbf{\Sigma}_i^{\text{2-way}}$ is:
\beq
\mathbf{\Sigma}_i^{\text{2-way}} \triangleq
         \left[
         \begin{array}{c}
             \left(T_i \sqrt{1 - T_i}\left( \nu-V_0 \right)\right) \mathbf{I}_2 \\
             \left(\sqrt{T_i\left(1 - T_i\right) \left(\nu^2 - 1\right)}\right) \mathbf{Z} \\
             -\sqrt{T_i\left(1-T_i\right)}\, \zeta_i\, \mathbf{Z}  \\
            \left(\sqrt{\left(1 - T_i\right) \left(\nu^2 - 1\right)}\right) \mathbf{I}_2
         \end{array}
         \right],
\label{eq53}
\eeq
where $\zeta_i \triangleq V_a+T_i\left(V_0-\nu\right)+\lambda_0+\sigma_{\text{g}}^2$.

The latter derivations can be used to calculate the symplectic eigenvalues of the covariance matrices and the conditional covariance matrix of Eve's modes, and consequently, yield the analytical expression (\ref{eq54}) for the effective SKR of the
two-way system (top of the next page).
\begin{figure*}[!t]
\beqarr
\text{SKR}_{\text{MIMO}}^{\text{2-way}}
= \sum_{i=1}^{r_H} \undb{E}_{T_i}
\left[ \text{SKR}_{i}^{\text{2-way}} \right]
\approx \sum_{i=1}^{r_H} \undb{E}_{T_i}
\left[ \beta \, I_{\text{B}} \left(a_i ; b_i^{\text{2-way}} \right)
- \sum_{q=1}^4 h_o \left( \lambda_{i_q}^{\text{2-way}} \right)
+ \sum_{q=5}^8 h_o \left( \lambda_{i_q}^{\text{2-way}} \right) \right]
\label{eq54}
\eeqarr
\noindent\rule{\textwidth}{.5pt}
\vspace{-0.8cm}
\end{figure*}
\subsection{Asymptotic Analysis}
We herein focus on the limit of high modulation, i.e., $V_s\gg V_0, v, \sigma_{\text{g}}^2$. For this case, the expression~\eqref{eq32} for the mutual information between Alice and Bob, when the one-way protocol is used, can be approximated as follows:
\begin{align}
& I_{\text{B}} \left(a_i; b_{i}^{\text{1-way}} \right) \nn \\
& \approx \sum_{k=0}^{\infty} \frac{e^{-\lambda_0} \lambda_0^k}{k!}
\left(-\log_2\left(\frac{e^{-\lambda_0} \lambda_0^k}{k!}\right)  + \frac{1}{2} \log_2 \left(2\pi e T_iV_s\right) \right. \nn \\
&+\log_2 \left(\sum_{\ell=0}^{R} \frac{e^{-\lambda_0} \lambda_0^{\ell}}
{\ell! 2\sqrt{\pi\left(T_i V_0 +\left(1-T_i\right)\nu+\sigma_{\text{g}}^2\right)}}
\right. \nn\\
& \qquad \left. \left. \times
\exp \left( - \frac{\left(k - \ell \right)^2}
{4\left(T_i V_0 +\left(1-T_i\right)\nu
+\sigma_{\text{g}}^2\right)}
\right) \right) \right).
\label{eq55}
\end{align}
In addition, the symplectic eigenvalues of Eve's covariance in ~\eqref{eq43} and the conditional covariance matrix in~\eqref{eq46}) can be approximated as follows:
\beqarr
&& \lambda_{i_1}^{\text{1-way}} \approx \nu, \quad \lambda_{i_2}^{\text{1-way}} \approx \left(1-T_i\right) V_s, \nn\\
&&\lambda_{i_3}^{\text{1-way}} \approx \nu, \quad \lambda_{i_4}^{\text{1-way}} \approx
\sqrt{\frac{\left(1 - T_i\right)V_s \nu}{T_i}}.
\label{eq56}
\eeqarr
Using the latter approximations along with the asymptotic expression of $h_0(x)\approx \log_2\left(\frac{e\,x}{2}\right)$ for $x\gg1$, the Holevo information in \eqref{eq41} can be also approximated as: 
\beqarr
\chi\left(e_i; b_i^{\text{1-way}}\right) \approx  \frac{1}{2}\,\log_{2}\!\left(\frac{T_i \left(1-T_i\right) V_s}{\nu} \right)\, .
\label{eq57}
\eeqarr
With the latter approximations and after performing some algebraic manipulations, an asymptotic expression for the SKR performance of the considered one-way MIMO FSO CV-QKD system is obtained as in \eqref{eq58} (top of the next page).
\begin{figure*}[!t]
\begin{align}
\text{SKR}_{\text{MIMO}}^{\text{1-way}}
& \approx \sum_{i=1}^{r_H} \undb{E}_{T_i}
\left[
\sum_{k=0}^{\infty}
\frac{e^{-\lambda_0} \lambda_0^k}{k!}
\left( -\log_2 \left(\frac{e^{-\lambda_0} \lambda_0^k}{k!}\right)+ \frac{1}{2} \log_2 \left(2 \pi e T_i^2 V_s\right) + \log_2 \left[\sum_{\ell=0}^{\infty}
\frac{e^{-\lambda_0} \lambda_0^{\ell}}
{\ell! 2\sqrt{\pi\left(T_iV_0+\left(1-T_i\right)\nu
+ \sigma_{\text{g}}^2\right)}} \right. \right. \right. \nn \\
& \qquad \qquad \qquad \left. \left. \left.
\times \exp \left(- \frac{\left(k -\ell \right)^2}
{4\left(T_iV_0+\left(1-T_i\right)\nu
+ \sigma_{\text{g}}^2\right)}  \right) \right] \right) - \frac{1}{2}\,\log_2 \!\left(\frac{T_i \left(1-T_i\right) V_s}{\nu} \right) \right]
\label{eq58}
\end{align}
\noindent\rule{\textwidth}{.5pt}
\vspace{-0.8cm}
\end{figure*}

Similarly, the expression~\eqref{eq39} for the mutual information between Alice and Bob for the case of the two-way protocol is approximated as follows:
\begin{align}
& I_{\text{B}} \left(a_i; b_{i}^{\text{2-way}} \right) \nn \\
& \approx \sum_{k=0}^{\infty} \frac{e^{-\lambda_0} \lambda_0^k}{k!}
\left(-\log_2\left(\frac{e^{-\lambda_0} \lambda_0^k}{k!}\right)  + \frac{1}{2} \log_2 \left(2\pi e T_iV_s\right) \right. \nn \\
&+\log_2 \left(\sum_{\ell=0}^{R} \frac{e^{-\lambda_0} \lambda_0^{\ell}}
{\ell! 2\sqrt{\pi\left(T_i^2 V_0 +\left(1-T_i^2\right)\nu+\left(1+T_i\right)\sigma_{\text{g}}^2\right)}}
\right. \nn
\end{align}
\begin{align}
& \qquad \left. \left. \times
\exp \left( - \frac{\left(k - \ell \right)^2}
{4\left(T_i^2 V_0 +\left(1-T_i^2\right)\nu
+\left(1+T_i\right)\sigma_{\text{g}}^2\right)}
\right) \right) \right).
\label{eq59}
\end{align}
Moreover, the symplectic eigenvalues of Eve's covariance matrix in~\eqref{eq50} and the conditional covariance matrix in~\eqref{eq52} are approximated as:
\begin{align}
& \lambda_{i_1}^{\text{2-way}} \approx \lambda_{i_2}^{\text{2-way}} \approx \nu, 
\quad \lambda_{i_3}^{\text{2-way}} \lambda_{i_4}^{\text{2-way}} \approx \left(1-T_i\right)^2 V_s^2, \nn\\
& \lambda_{i_5}^{\text{2-way}} \approx \nu,
\quad \lambda_{i_6}^{\text{2-way}} \approx
\sqrt{\frac{\left(\left( 1 + T_i^3 \right)\nu + \left( 1 - T_i \right) T_i^2 V_0 \nu^2 \right)}
{\left(1-T_i\right)T_i^2V_0+\left(1+T_i^3\right)\nu}}, \nn \\
& \lambda_{i_7}^{\text{2-way}} \lambda_{i_8}^{\text{2-way}} \!\approx \!  
\sqrt{\frac{V_s^3 \left(1 - T_i \right)^3\left(\left(1-T_i\right)T_i^2V_0+\left(1+T_i^3\right)\nu\right)}{T_i}},
\label{eq60}
\end{align}
yielding the following approximation for the Holevo information in \eqref{eq49}: 
\beqarr
\chi\left(e_i; b_i^{\text{2-way}}\right) \!\!\!\!\!\!\!\!\!\!&&\approx  \frac{1}{2}\,\log_{2}\!\left(\frac{T_i \left(1-T_i\right) V_s}{T_i^2V_0+\nu+T_i^3\left(\nu-V_0 \right)} \right)\nn\\
&&\quad+ h_0\left(\nu\right) -h_o\left(\lambda_{i_6}^{\text{2-way}} \right).
\label{eq61}
\eeqarr
Using all above, an asymptotic expression for the SKR
performance of the considered two-way MIMO FSO CV-QKD
system is given as in~\eqref{eq62} (top of the next page).
\begin{figure*}[!t]
\beqarr
&&\text{SKR}_{\text{MIMO}}^{\text{2-way}}
\approx \sum_{i=1}^{r_H} \undb{E}_{T_i}
\left[
\sum_{k=0}^{\infty}
\frac{e^{-\lambda_0} \lambda_0^k}{k!}
\left( -\log_2\left(\frac{e^{-\lambda_0} \lambda_0^k}{k!}\right)
+ \log_2 \left[\sum_{\ell=0}^{\infty}
\frac{e^{-\lambda_0} \lambda_0^{\ell} \times \exp \left(- \frac{\left(k -\ell \right)^2}
{4\left(T_i^2 V_0 +\left(1-T_i^2\right)\nu+\left(1+T_i\right)\sigma_{\text{g}}^2\right)}  \right)}
{\ell! 2\sqrt{\pi\left(T_i^2 V_0 +\left(1-T_i^2\right)\nu+\left(1+T_i\right)\sigma_{\text{g}}^2\right)}} 
  \right]  \right. \right. \nn \\
&& \qquad \qquad \qquad \qquad \quad\left. \left.
 + \frac{1}{2} \log_2 \left(2 \pi e T_i^2 V_s\right)\right)
 -\frac{1}{2}\,\log_{2}\!\left(\frac{T_i \left(1-T_i\right) V_s}{T_i^2V_0+\nu+T_i^3\left(\nu-V_o \right)} \right)
 - h_0\left(\nu\right) +h_o\left(\lambda_{i_6}^{\text{2-way}} \right) \right]
\label{eq62}
\eeqarr
\noindent\rule{\textwidth}{.5pt}
\vspace{-0.8cm}
\end{figure*}

Without loss of generality, we also assert the importance of investigating a scenario where Eve's variance is precisely set to $\nu = 1$ and the thermal noise variance $V_0$ closely aligns with  $\nu$, i.e., $V_0 \approx \nu$ \cite{weedbrook2010quantum, ottaviani2020terahertz}. This consideration is pivotal for a comprehensive understanding of the system's dynamics when variances are comparable. 
Using \eqref{eq58} and \eqref{eq62}, the difference between the SKRs obtained for the one- and two-way MIMO FSO systems for CV-QKD can be computed as in \eqref{eq63} (top of the next page).
\begin{figure*}[!t]
\begin{align}
\Delta\text{SKR}_{\text{MIMO}}
& = \text{SKR}_{\text{MIMO}}^{\text{2-way}}
-\text{SKR}_{\text{MIMO}}^{\text{1-way}}\nn\\
& \approx \sum_{i=1}^{r_H} \undb{E}_{T_i}
\left[
\sum_{k=0}^{\infty}
\frac{e^{-\lambda_0} \lambda_0^k}{k!}
\left( \log_2 \left[ 
\frac{\left(1  + \sigma_{\text{g}}^2 \right)
\sum\limits_{\ell=0}^{\infty}
\frac{e^{-\lambda_0} \lambda_0^{\ell}}
{\ell!} \exp \left(\frac{-\left(k-\ell\right)^2}{4\left(1+\left(1+T_i\right)\sigma_{\text{g}}^2\right)} \right) }
{\left(1+\left(1+T_i\right)\sigma_{\text{g}}^2 \right)
\sum\limits_{\ell=0}^{\infty}
\frac{e^{-\lambda_0} \lambda_0^{\ell}}
{\ell !} \exp \left(\frac{-\left(k-\ell\right)^2}{4\left(1+\sigma_{\text{g}}^2\right)} \right)}\right]\right)+\frac{1}{2}
\log_2 \left(1+T_i^2V_0 \right) \right]
\label{eq63}
\end{align}
\noindent\rule{\textwidth}{.5pt}
\vspace{-0.8cm}
\end{figure*}
Furthermore, for $T_i \ll 1$ $\forall i=1,\ldots, r_H$, the differential SKR in \eqref{eq63} can be further simplified as:
\beq
\Delta\text{SKR}_{\text{MIMO} \big|
\left\{T_i \right\}_{i=1}^{r_H} \ll 1}
\approx \frac{1}{2} \sum_{i=1}^{r_H} \undb{E}_{T_i}
\left[ \log_2 \left(1 + T_i^2 V_0 \right) \right].
\label{eq64}
\eeq

It is noted that $T_i$'s can be considered as statistically independent. Thus, the differential gain obtained by using the two-way protocol over the one-way one is proportional to $r_H$. This indicates that increasing the MIMO configuration results in SKR gains for the considered FSO CV-QKD system.
\begin{figure*}[!t]
    \centering
    \includegraphics[width=18cm,height=6cm]{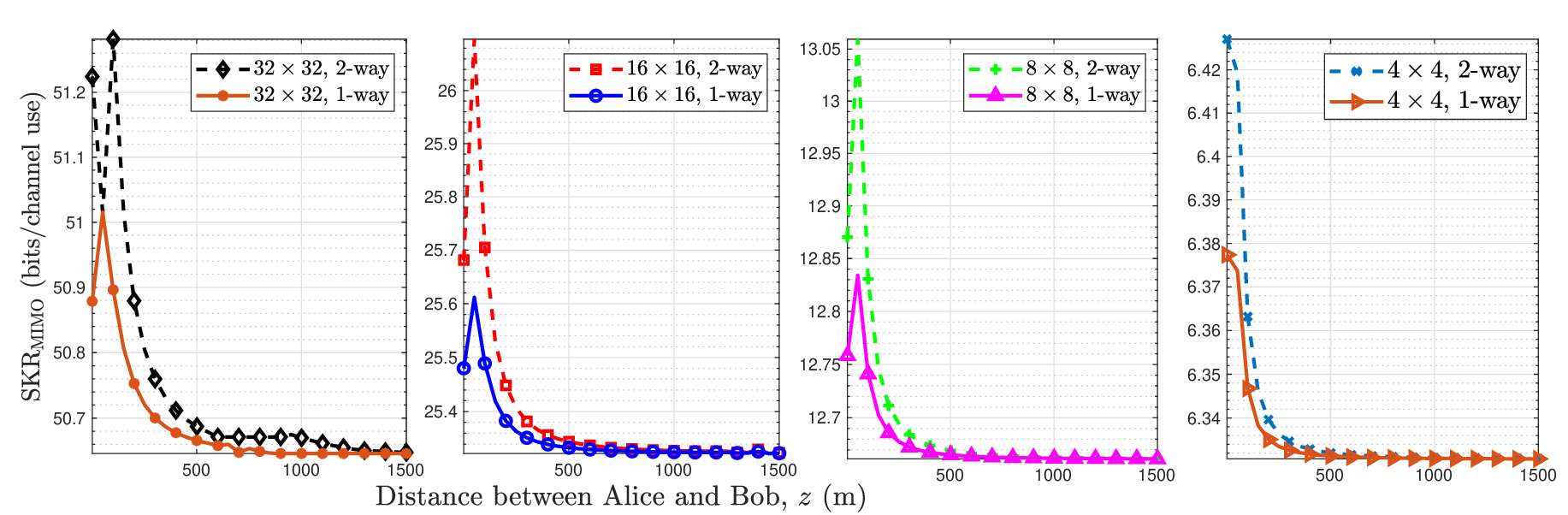}
    \caption{$\text{SKR}_{\text{MIMO}}$ vs. distance between Alice and Bob, $z$, for the MIMO configuration $N_T=N_R=\{4, 8, 16, 32\}$, $\lambda_0 = 1$, $\sigma_{\text{g}}^2 = 0.001$, $\eta=1, \beta = 1$, and $C_n^2 = 10^{-15} \text{m}^{-2/3}$.}
    \label{f4}
\end{figure*}
\section{Numerical Results and Discussion}
This section presents the numerical results that corroborate the analytical framework presented in this paper.
We have considered the following system parameters for carrying out the numerical investigation: $N_T=N_R=r_H$, $\lambda = 1550 \, \text{nm}$, $ w= 2.5 \, \text{mm}$, $a_r = 10 \, \text{cm}$, $c=3 \times 10^8 \, \text{m/s}$, $T_o=296 \, \text{K}$, $h=6.63 \times 10^{-34} \, \text{Js}$, $k_B=1.38 \times 10^{-23} \, \text{J/K}$, $\nu=1$, and $V_s=10^3$. Further, $V_0 =2\bar{n} + 1$, where $\bar{n} = 1 / (\exp\left(h f_c/k_B T_o \right) - 1)$, $f_c = c / \lambda $, and 
$\delta =0.43\times10^{-3} \, \text{dB/m}$.

Figure~\ref{f4} presents the plots of the $\text{SKR}_{\text{MIMO}}$ for both the one- and two-way protocols used with the considered MIMO FSO CV-QKD system under hybrid quantum noise as a function of the transmission distance $z$ between Alice and Bob. It is observed that, across all MIMO configurations, the SKR exhibits a sharp decay over short ranges, eventually transitioning into a shallow, distance-insensitive tail. It is also shown that the two-way protocol consistently outperforms the one-way one, with the relative advantage being most notable at short ranges and gradually narrowing as the distance $z$ increases. As the MIMO order increases, the performance curves rise across all values of $z$. This trend remains consistent over varying distances, although the added benefits of larger MIMO tend to diminish as the distance increases. This phenomenon can be attributed to the dominance of path loss and pointing error due to turbulence at longer transmission distances, and to the more performance degradation from hybrid quantum noise in larger MIMO configurations. 

It has been observed that the analytical bounds presented in \cite{mouli_hybridnoise_Jan2025} closely follow the the numerically evaluated classical mutual information which leads to the same with SKR. This finding indicates a close tightness of the bounds for the considered system model. Finally, it can be seen in Fig.~\ref{f4} that, within the considered range of transmission distances and MIMO configurations, the analytical bounds reported in \cite{mouli_hybridnoise_Jan2025} closely follow the numerically evaluated SKRs, indicating near-tightness of the bounds in this regime.

\begin{figure}[!t]
    \centering
    \includegraphics[width=8cm,height=7cm]{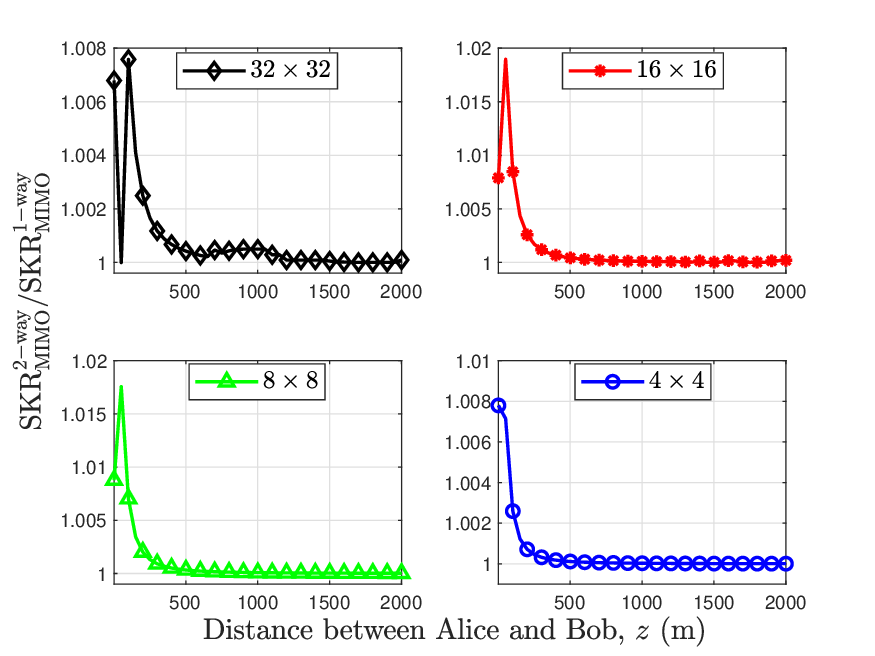}
    \caption{SKR ratio vs. distance between Alice and Bob, $z$, for $N_T=N_R=\{4, 8, 16, 32\}$ MIMO configuration, $\lambda_0 = 1$, $\sigma_{\text{g}}^2 = 0.001$, $\eta=1, \beta = 1$, and $C_n^2 = 10^{-15} \text{m}^{-2/3}$.}
    \label{f5}
\end{figure}
\begin{figure}[!t]
    \centering
    \includegraphics[width=8cm,height=7cm]{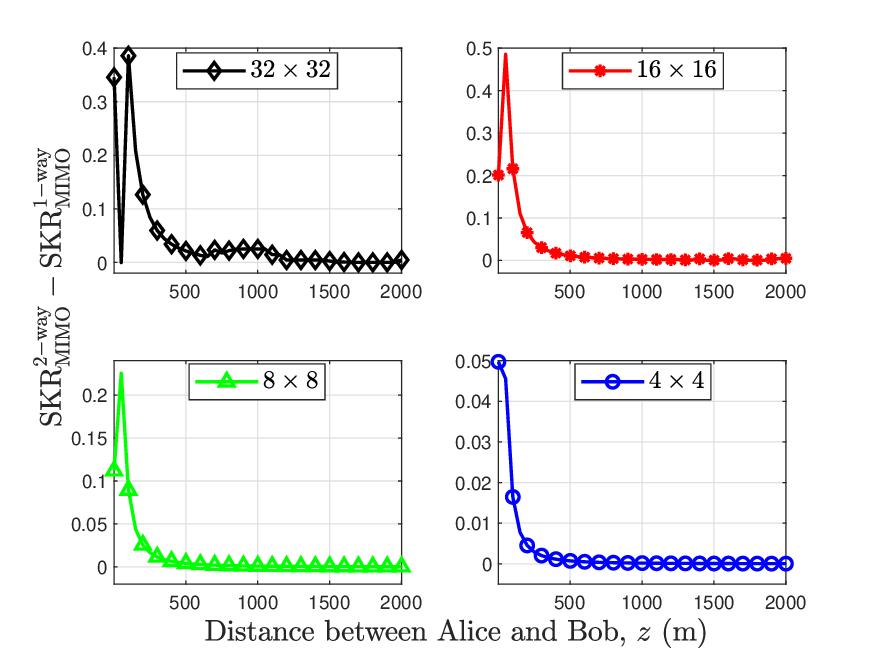}
    \caption{SKR difference vs. distance between Alice and Bob, $z$, for $N_T=N_R=\{4, 8, 16, 32\}$, $\lambda_0 = 1$, $\sigma_{\text{g}}^2 = 0.001$, $\eta=1, \beta = 1$, and $C_n^2 = 10^{-15} \text{m}^{-2/3}$. }
    \label{f6}
\end{figure}
Figure~\ref{f5} illustrates the ratio of the SKR between the two- and one-way protocols for varying MIMO configurations. It can be observed that, across all MIMO sizes, the ratio exceeds unity at a short range, indicating a distinct two-way advantage. However, as the distance increases, this ratio quickly diminishes toward unity. Nevertheless, the SKR ratio remains above unity over a longer transmission distance compared to lower MIMO configurations, indicating a monotonic increase in achievable range as the MIMO order increases. It is also depicted that the peak SKR ratio is lower for both very high and very low MIMO configurations than for intermediate ones. This behavior arises because higher-order MIMO systems are more susceptible to inter-channel interference at the receiver, whereas lower-order MIMO systems are less effective at mitigating fading losses and noise effects. Moreover, Fig.~\ref{f6} depicts the difference in SKRs between the two- and one-way protocols with respect to the distance between Alice and Bob. A similar trend to the SKR ratio is demonstrated, with the SKR difference falling quickly at lower distances and remaining marginally greater than zero at longer distances.

\begin{figure}[!t]
    \centering
    \includegraphics[width=8cm,height=7cm]{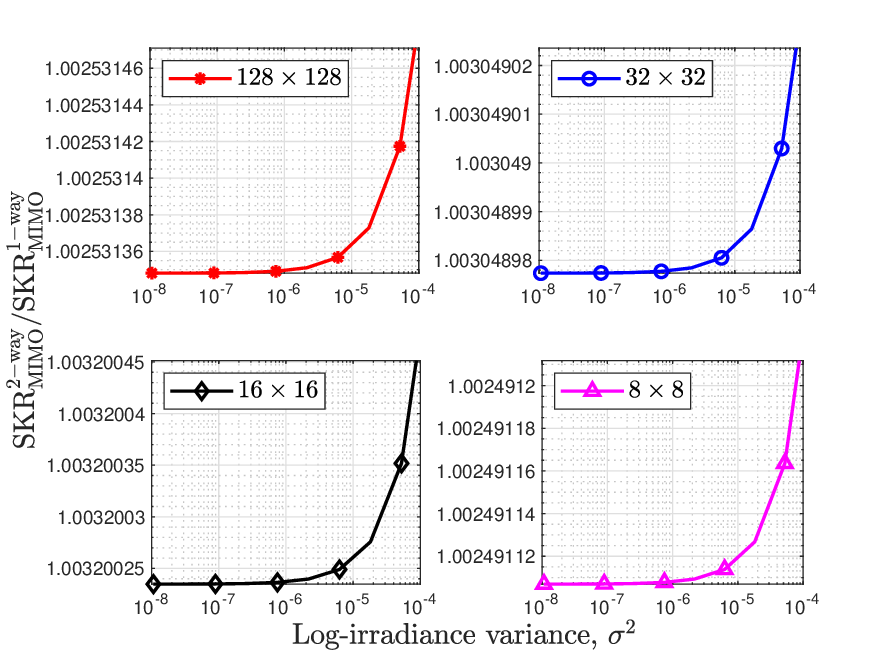}
    \caption{SKR ratio vs. lognormal irradiance variance, $\sigma^2$, for $N_T=N_R=\{8, 16, 32, 128\}$, $\eta=1$, $z=200$ m, and $C_n^2=\left(10^{-17} -10^{-14}\right) \text{m}^{-2/3}$.}
    \label{f7}
\end{figure}
Figure~\ref{f7} illustrates the relative benefits of the two-way protocol under turbulence via the quantity $\text{SKR}_{\text{MIMO}}^{\text{2-way}}
-\text{SKR}_{\text{MIMO}}^{\text{1-way}}$, as a function of the log-irradiance variance $\sigma^{2}$. It is shown that, for all MIMO setups, the ratio remains above unity for the entire range of $\sigma^2$. The curves show a consistent and convex rise, particularly as the scintillation as $\sigma^2\geq 10^{-6}$. This indicates that the effect of atmospheric scintillation is not in the same proportion for both protocols, showing that the two-way system is resilient to atmospheric scintillation.

\begin{figure}[!t]
    \centering
    \includegraphics[width=8cm,height=7cm]{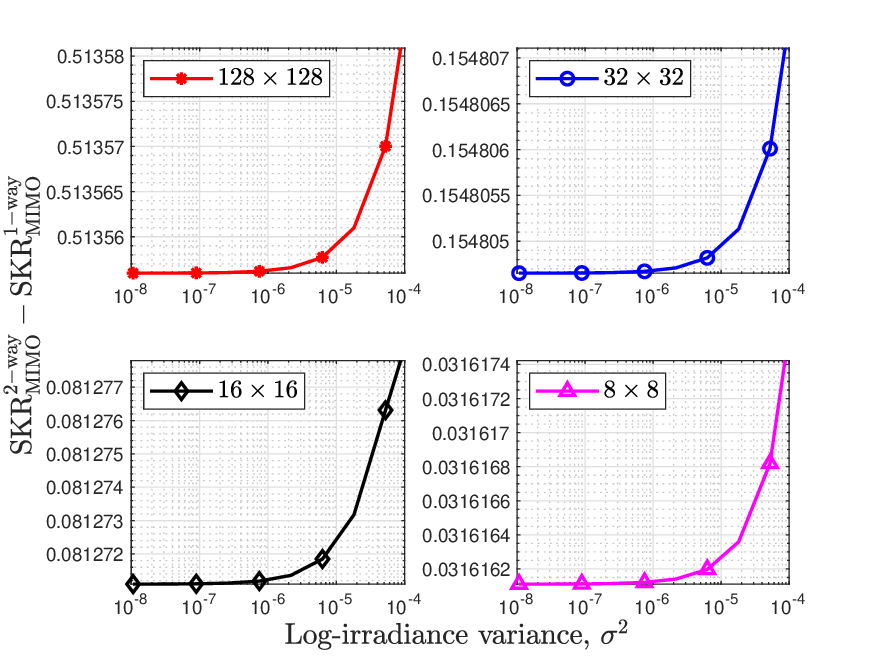}
    \caption{SKR difference v.s lognormal irradiance variance, $\sigma^2$, for $N_T=N_R=\{8, 16, 32, 128\}$, $\eta=1$, $z=200$ m, and $C_n^2=\left(10^{-17} -10^{-14}\right) \text{m}^{-2/3}$.}
    \label{f8}
\end{figure}
Figure~\ref{f8} illustrates the differential SKR gain of the two-way protocol over the one-way counterpart, $\Delta\text{SKR}_{\text{MIMO}}$, for different $\sigma^2$ values reflecting the strength of atmospheric scintillation. The results show that $\Delta\text{SKR}_{\text{MIMO}}$ increases as the MIMO configuration scales up. With respect to $\sigma^2$, this metric remains nearly constant at first, begins to rise when $\sigma^2 \geq 10^{-6}$, and then exhibits only a marginal increase at higher values of $\sigma^2$. This again indicates that the overall impact of atmospheric scintillation on the two-way protocol is weaker than on the one-way protocol.

\begin{figure}[!t]
    \centering
    \includegraphics[width=8.7cm,height=6.5cm]{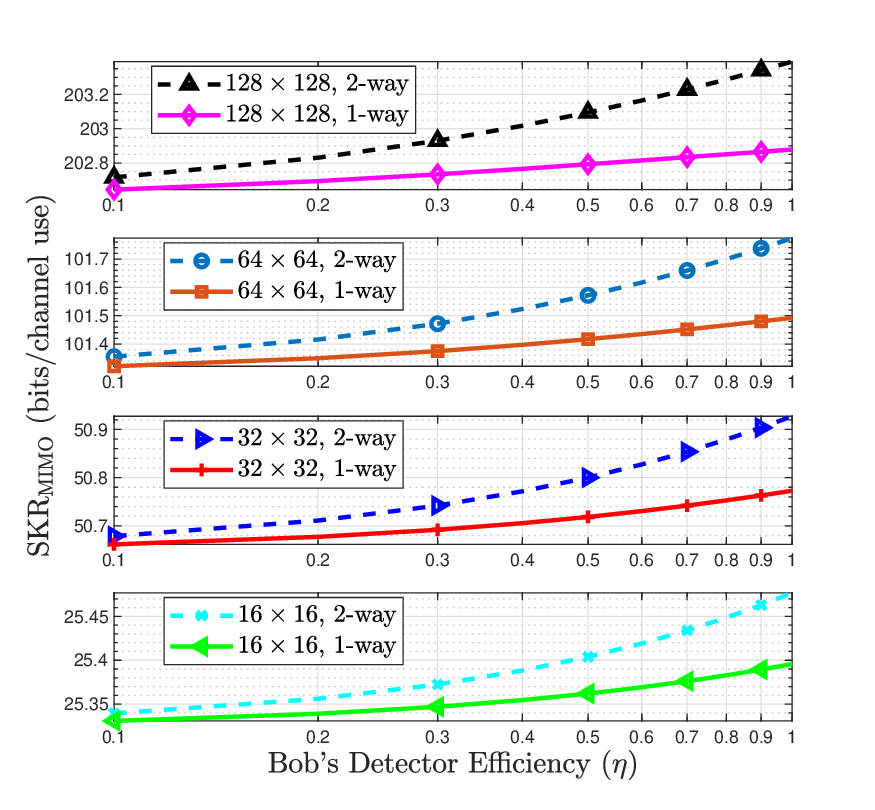}
    \caption{SKR vs. detector efficiency $\eta$ for $N_T=N_R=\{16, 32, 64, 128\}$ at $z=200$ m for hybrid quantum noise parameter $\lambda_0 = 1$, $\sigma_{\text{g}}^2 = 0.001$, and $C_n^2=10^{-15}\, \text{m}^{-2/3}$.}
    \label{f9}
\end{figure}
The metric $\text{SKR}_{\text{MIMO}}$ as a function of Bob's detector efficiency $\eta$ for $N_R=N_T=\{16,32,64,128\}$ is plotted in Fig.~\ref{f9}. Each subplot compares the two- and one-way protocols under identical parameter settings. It is demonstrated that, for each MIMO configuration, $\text{SKR}_{\text{MIMO}}$ exhibits a consistent increase as $\eta$ rises. Throughout the entire range of $\eta$, the two-way protocol performs better than the one-way, with the most noticeable difference occurring at moderate to high efficiency values, i.e, $\eta \geq 0.5$. It can be clearly observed from these plots that the SKR values almost double with a doubling in the MIMO configuration, thus, justifying the result obtained in \eqref{eq64}. Furthermore, the sensitivity of the system's SKR to $\eta$ in terms of the value of $\frac{\text{d}\,\text{SKR}_{\text{MIMO}}}{\text{d}\eta}$ increases with the MIMO order. The gains from increasing $\eta$ are modest at lower MIMO setups but become more significant for the higher ones.

\begin{figure}[!t]
    \centering
    \includegraphics[width=8.7cm,height=6.5cm]{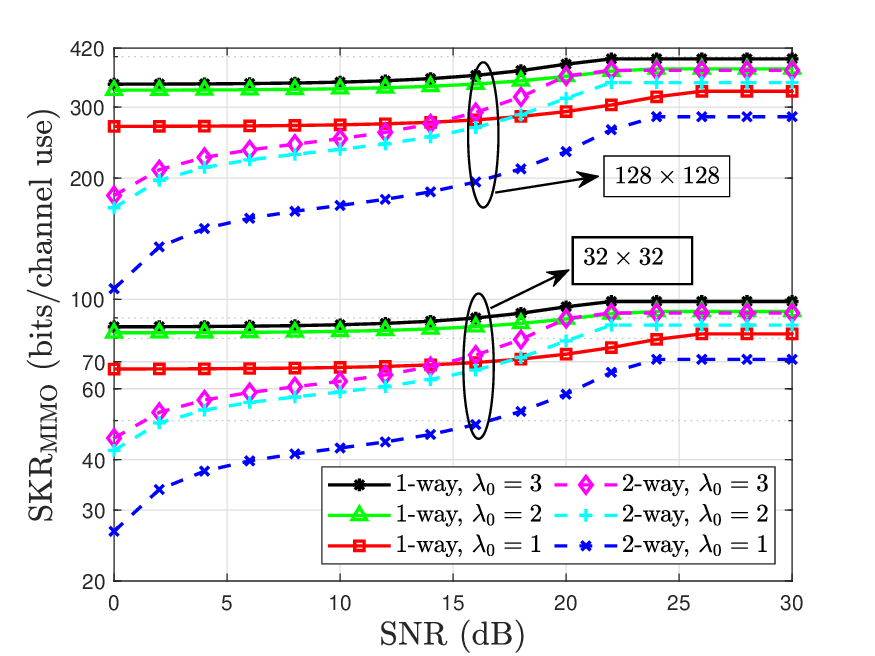}
    \caption{$\text{SKR}_{\text{MIMO}}$ vs. \text{SNR} (dB) for $N_R=N_T=\{32,128\}$, $z=500$ m, $\lambda_0 =\{1,2,3\}$, and $T_i = 0.5$.}
    \label{f10}
\end{figure}
Finally, Fig.~\ref{f10} illustrates $\text{SKR}_{\text{MIMO}}$ as a function of $\text{SNR}$ for both protocols. It is shown that, in the low $\text{SNR}$ range, the curves obtained from the two-way protocol fall below the ones obtained from the one-way counterpart. This is attributed to the additive noise, which has a greater impact in the two-way protocol. Furthermore, the two-way protocol demonstrates stronger sensitivity to $\text{SNR}$, exhibiting a steeper increase compared to one-way systems. In addition, it can be seen that SKR for both protocols saturates at higher $\text{SNR}$ values. Higher MIMO system configurations lead to a notable SKR improvement,  with $\lambda_0$ values consistently increasing across the entire $\text{SNR}$ range, as compared to lower MIMO setups.

\section{Conclusions}
This paper presented a MIMO FSO system with two legitimate users, employing CV-QKD for secret key exchange in the presence of an eavesdropper, attempting to jeopardize secure communications through a collective Gaussian attack. The considered FSO channels were subjected to atmospheric turbulence, introducing beam spreading and wandering, pointing error, attenuation, and turbulence-induced fading. The legitimate system was also assumed to be subjected to hybrid quantum noise, degrading CV-QKD performance. A mathematical framework for characterizing the transmissivity of the considered FSO MIMO channel was presented. Furthermore, two protocols, a one-way and a two-way, were proposed for secret key exchange. Owing to the statistics of the hybrid quantum noise, bounds on the mutual information between the coherent states of the eavesdropper and the legitimate receiver were computed for both protocols. In addition, for the case where the legitimate receiver employs homodyne detection and RR, closed-form and asymptotic expressions for the system's SKR for both protocols were derived. Our numerical investigations showcased that the differential gain obtained by employing the two-way protocol over the one-way counterpart is proportional to the rank of the MIMO FSO channel. Increasing the MIMO configuration while using the two-way protocol resulted in improved SKR performance, facilitating secure secret key exchange over larger distances between the legitimate ends.

\bibliographystyle{IEEEtran}
\bibliography{IEEEabrv,bibliography}
\end{document}